%% file: main.tex
\documentclass[conference]{IEEEtran}
\IEEEoverridecommandlockouts
\usepackage{cite}
\usepackage{amsmath,amssymb,amsfonts}
\usepackage{balance}
\usepackage{graphicx}
\usepackage{boldline} 
\usepackage{multirow}
\usepackage{booktabs}
\usepackage{textcomp}
\usepackage{xcolor}
\def\BibTeX{{\rm B\kern-.05em{\sc i\kern-.025em b}\kern-.08em
    T\kern-.1667em\lower.7ex\hbox{E}\kern-.125emX}}
\usepackage[noend]{algpseudocode}
\usepackage[caption=false]{subfig}
\usepackage{xspace}
\newcommand{\PaperAcronym}{AutoCure\xspace}
\newcommand{\quotes}[1]{``#1''}
\algnewcommand\Or{\textbf{or}}
\algnewcommand\And{\textbf{and}}
\newcommand{\ra}[1]{\renewcommand{\arraystretch}{#1}} 

\begin{document}

\title{\PaperAcronym: Automated Tabular \\Data Curation Technique for ML Pipelines
}

\author{\IEEEauthorblockN{Mohamed Abdelaal}
\IEEEauthorblockA{\textit{Software AG} \\
Darmstadt, Germany \\
mohamed.abdelaal@softwareag.com}
\and
\IEEEauthorblockN{Rashmi Koparde}
\IEEEauthorblockA{\textit{Otto von Guericke University Magdeburg} \\
Magdeburg, Germany \\
rashmi.koparde@st.ovgu.de}
\and
\IEEEauthorblockN{Harald Schöning}
\IEEEauthorblockA{\textit{Software AG} \\
Darmstadt, Germany \\
harald.schoening@softwareag.com}
}

\maketitle

\begin{abstract}
Machine learning algorithms have become increasingly prevalent in multiple domains, such as autonomous driving, healthcare, and finance. In such domains, data preparation remains a significant challenge in developing accurate models, requiring significant expertise and time investment to search the huge search space of well-suited data curation and transformation tools. To address this challenge, we present \PaperAcronym, a novel and configuration-free data curation pipeline that improves the quality of tabular data. Unlike traditional data curation methods, \PaperAcronym synthetically enhances the density of the clean data fraction through an adaptive ensemble-based error detection method and a data augmentation module. In practice, \PaperAcronym can be integrated with open source tools, e.g., Auto-sklearn, H2O, and TPOT, to promote the democratization of machine learning. As a proof of concept, we provide a comparative evaluation of \PaperAcronym against 28 combinations of traditional data curation tools, demonstrating superior performance and predictive accuracy without user intervention. Our evaluation shows that \PaperAcronym is an effective approach to automating data preparation and improving the accuracy of machine learning models.
\end{abstract}


\input{sections/introduction.tex}
\input{sections/concept.tex}
\input{sections/error_detection.tex}
\input{sections/augmentation.tex}

\input{sections/evaluation.tex}
\input{sections/related_work.tex}
\input{sections/conclusion.tex}

\section*{Acknowledgment}
This work was supported (in part) by the Federal Ministry of Education and Research through grants 02L19C155, 01IS21021A (ITEA project number 20219).

\balance
\bibliographystyle{plain}
\bibliography{references}

\end{document}

%% file: sections/introduction.tex
\section{Introduction}
\label{sec:introduction}
%
%
\textbf{Data Quality Problems:} In the recent decades, machine learning (ML) has developed a strong impact in several application domains, including autonomous driving, gaming, healthcare, logistics, and finance \cite{dong2021survey}. Developing ML models, in such domains, typically involves data acquisition from multiple distinct sources. For instance, perception and state estimation systems in self-driving cars usually employ ML models trained on data collected from sensor readings, GPS fixes, and historical records \cite{yeong2021sensor}. After data acquisition, data preparation and transformation processes are commonly carried out to bring the collected data into a state suitable for model building. In fact, the performance of such ML models broadly depends on the quality of the input data \cite{neutatz2021cleaning}. Specifically, the predictive performance may drastically degrade whenever the input data is noisy or contains erroneous instances. In general, real-world data mostly suffers from heterogeneous error profiles due to improper join operations, noisy communication channels, inaccurate and/or incomplete manual data entry, etc. \cite{ilyasdata}. Due to such technical/human-related problems, different error types, e.g., outliers, pattern/rule violations, duplicates, inconsistencies, and implicit/explicit missing values, may simultaneously emerge in a data set.

%
\textbf{Challenges:} Since high data quality is a necessary prerequisite for improving the performance of ML models, the input data has to be properly curated before being employed for modeling tasks. A typical data curation pipeline usually begins with detecting erroneous instances before repairing them. In fact, there exist plenty of commercial and open source tools for error detection and repair, e.g., Talend, OpenRefine, Trifacta, Tamr, NADEEF, and HoloClean \cite{max_min16}. However, such tools still suffer from several problems. First, they usually require domain knowledge and skilled individuals who can formulate such knowledge as a set of rules/constraints. Exceptions to these rules/constraints are typically reported to the users to take appropriate corrective actions, either by correcting the data, or by fine-tuning the data and/or the rule definitions. Due to user involvement, it may become challenging to adopt these data curation tools without users possessing sufficient data expertise. A possible solution could be the adoption of an automated rule generation tool, e.g., Metanome \cite{metanome15}, DCFinder \cite{dcfinder19}, and RTClean \cite{rtclean23}. Nevertheless, the performance of such tools, in terms of the quality of the generated rules, varies broadly for different datasets, according to our experiments \cite{rein23}. Moreover, their high computational complexity hinders their integration with data curation tools.

Second, most ML-based error detection tools, e.g., RAHA \cite{raha19} and HoloDetect \cite{holodetect19}, cannot recognize the type of the detected errors, i.e., whether they are rule/pattern violations, outliers, missing values, or duplicates. They simply train a detection classifier which differentiates between erroneous and clean data instances. Due to lack of knowledge about the detected error types, the task of selecting a well-suited data repair tool becomes non-trivial. As a workaround, one may implement multiple data repair tools. However, this solution may increase the complexity of the curation pipeline, since the search space of repair candidates drastically increases. For example, if a data instance $X_i=x_{i,1},\cdots,x_{i,k}$ has an empty cell $x_{i,j}=NaN$, where $k$ is the number of columns, the cell $x_{i,j}$ has to be properly imputed. Knowing that there exist plenty of imputation methods which generate distinct values $a_1,\cdots, a_n$, it becomes challenging to select the optimal repair candidate. To this end, it becomes necessary to further extend the data preparation pipeline, which in turn increases the complexity, via implementing additional tools, e.g., BoostClean \cite{boostclean17} and CPClean \cite{cpclean20}.

%
\textbf{Proposed Method:} To overcome these challenges, we introduce a novel automated data curation pipeline, referred to as \PaperAcronym, which comprises two main modules, namely the \textit{adaptive ensemble-based error detection} module and the \textit{clean data augmentation} module. The core idea behind \PaperAcronym is to synthetically enhance the density of the clean fraction of the input data (which typically consists of clean and dirty data instances). According to the information theory, the process of noise reduction is equivalent to adding up more data with similar quality \cite{shannon2001}. Pillared on this theory, \PaperAcronym strives to increase the proportion of clean data instances for the sake of reducing the impact of noisy/erroneous data instances. Accordingly, we can avoid the problems of data repair via replacing it with a clean data augmentation module.

\begin{figure}[t]
	\centering
	\subfloat[Impact of noise on model building]{\label{fig:a}\includegraphics[width=0.85\columnwidth]{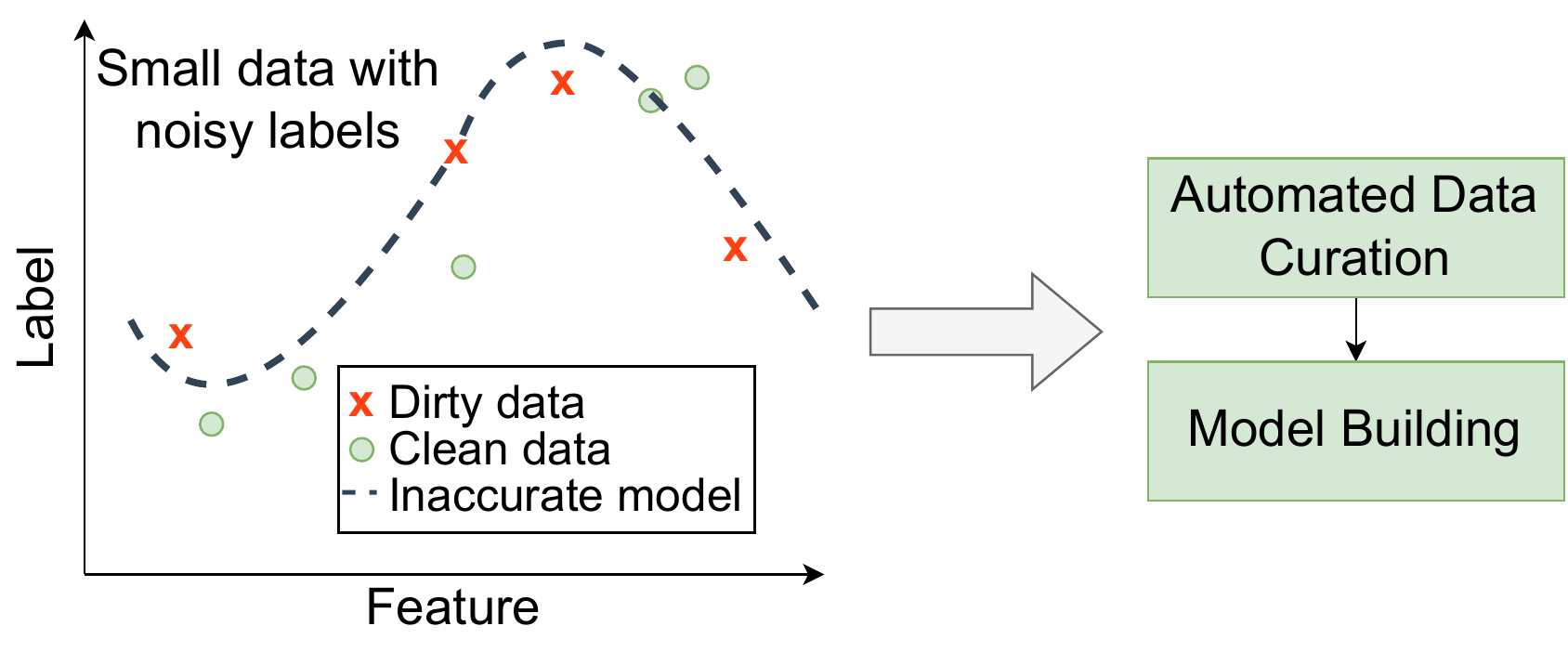}} \hfill \hspace{2cm}
	\subfloat[Traditional Curation]{\label{fig:b}\includegraphics[width=0.5\columnwidth]{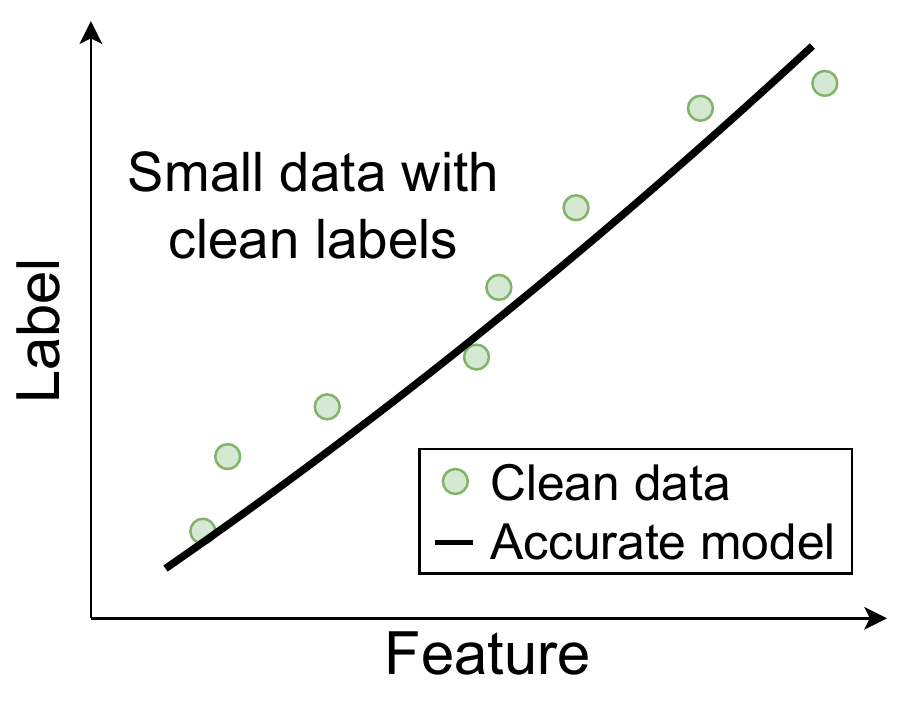}} \hfill
	\subfloat[\PaperAcronym]{\label{fig:c}\includegraphics[width=0.5\columnwidth]{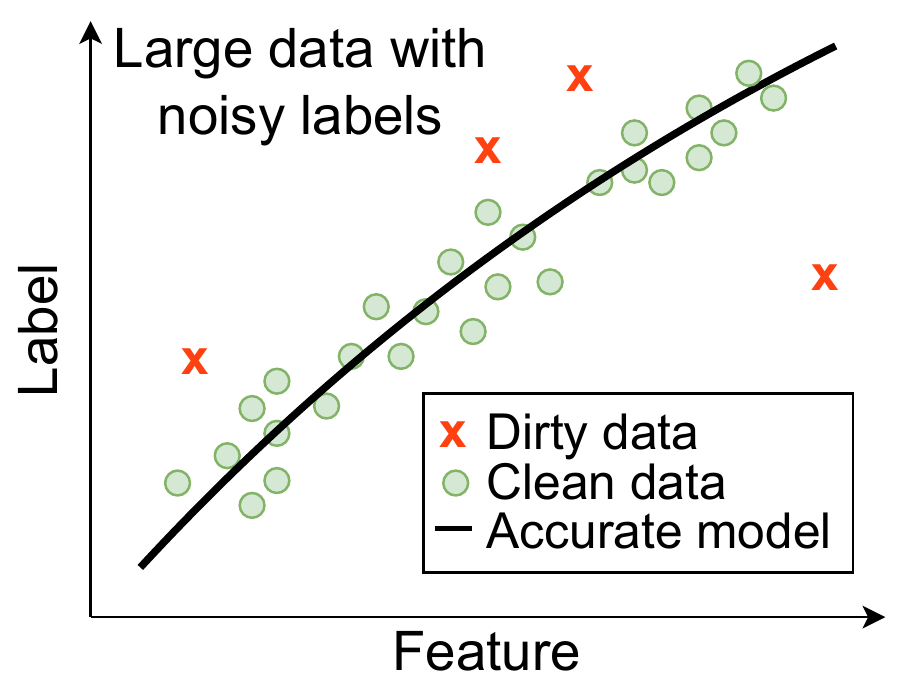}} \hfill
	\caption{Three ML models trained on different types of data}
	\label{fig:informationtheory}
\end{figure}

To further clarify the idea behind \PaperAcronym, Figure~\ref{fig:informationtheory} shows an illustrative example of three different scenarios. Figure~\ref{fig:a} shows an ML model (dashed curve) trained on a small data set, which includes a set of clean instances (green circles) and a set of noisy data instances (red circles). Due to the noisy instances, such an ML model fail to precisely describe the linear data set. To improve the predictive performance, we may repair the noisy data instances using a traditional data repair method, as depicted in Figure~\ref{fig:b}. However, the repair process is usually complex, involving user intervention and additional steps to select the best repair candidates. Moreover, traditional data repair methods can not restore the true values of the noisy instances. A second strategy, adopted by \PaperAcronym, for dealing with the poor data quality problem is to deliberately increase the density of clean instances, as Figure~\ref{fig:c} depicts. In this case, the impact of the dirty instances on the ML modeling process is significantly reduced. As a result, the generated ML model broadly resembles the curve shown in Figure~\ref{fig:b}.

\textbf{Summary of Contributions:} To sum up, the paper provides the following contributions: (1) We introduce a novel two-stages curation pipeline for dealing with erroneous tabular data through increasing the density of clean data instances. (2) We propose an adaptive ensemble-based error detection method. The proposed detector annotates a cell $x_{i,j}$ as dirty, if the cell is detected by at least $k$ base detectors. \PaperAcronym dynamically adapts the threshold $k$ to overcome the \textit{data exclusion} problems which typically occur while extracting clean data instances from the input data set. Such data problems usually happen due to inaccurate detections of the base detectors (cf. Section~\ref{sec:adaptive_detector}). (3) We conduct extensive experiments to evaluate \PaperAcronym relative to the ground truth, to obtain the performance upper-bound, and a wide collection of baseline methods in terms of the modeling accuracy and the training time. The results show that \PaperAcronym constantly achieves a comparable performance to the ground truth data together with requiring less training time compared to the case of running several repair methods to select the best repair candidates. To the best of our knowledge, \PaperAcronym is the first work which automates data curation in ML pipelines via reducing the impact of erroneous data and considering the density of the clean fraction as a practical solution for the data quality problems. 

\textbf{Structure of the Paper:} The remainder of this paper is structured as follows. Section~\ref{sec:concept} introduces the different components in the proposed pipeline, together with highlighting the main assumptions. Section~\ref{sec:adaptive_detector} presents the adaptive ensemble-based error detection method and explains the data exclusion problems. In Section~\ref{sec:augmentation}, we discuss the augmentation of clean data using the so-called variational autoencoder (VAE). Section~\ref{sec:evaluation} introduces our proof-of-concept implementation, before presenting the obtained results for different data sets. Section~\ref{sec:related_work} discusses the related work, before Section~\ref{sec:conclusion} concludes the paper with an outlook on future work.

%% file: sections/concept.tex
\section{Overview \& Architecture}
\label{sec:concept}

\begin{figure*}[tbph]
	\centering
	\includegraphics[width=0.95\linewidth]{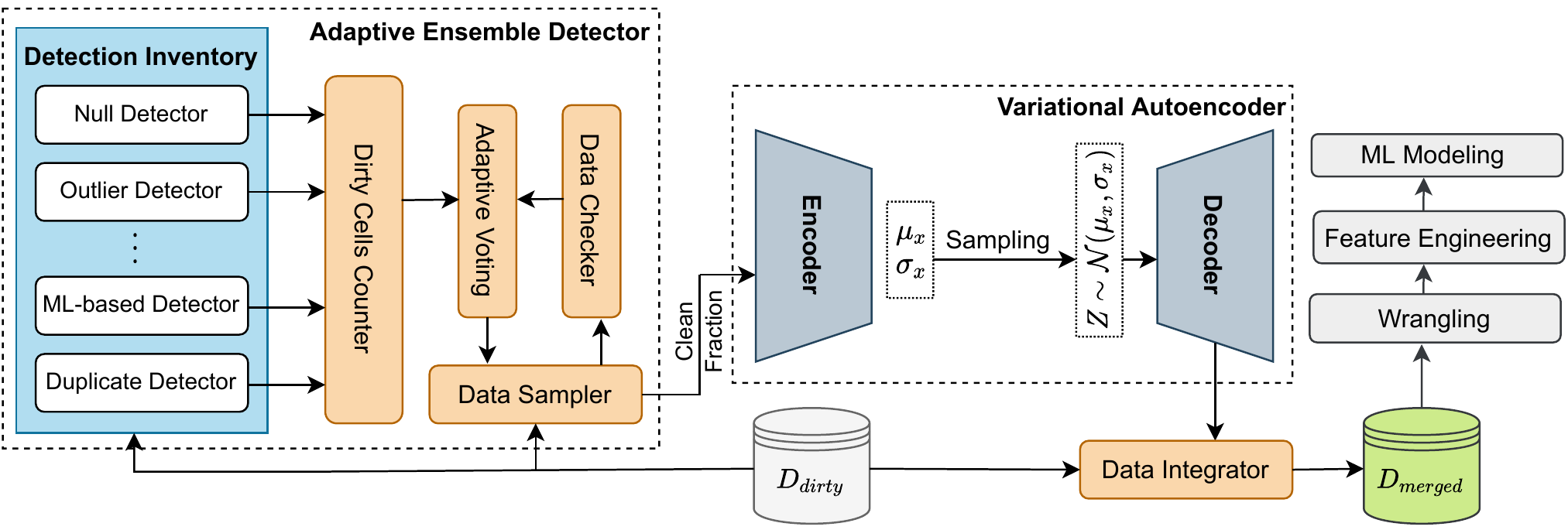}
	\caption{Architecture of the proposed method}
	\label{fig:architecture}\vspace{-0.5cm}
\end{figure*}

In this section, we present the architecture of \PaperAcronym together with our assumptions. The proposed architecture consists of two modules, an adaptive ensemble-based error detector and a clean data augmentation, which perform three main tasks, including (1) the detection of erroneous data instances, (2) the extraction of clean data fraction including all classes, and (3) the generation of additional data from the same distribution as the clean fraction. Figure~\ref{fig:architecture} gives an overview over all components with their respective inputs. At the outset, a dirty data set is used as an input to the adaptive ensemble-based error detection module. The rationale behind such a module is to maximize the detection recall, defined as the fraction of erroneous data instances that are detected. Moreover, such an adaptive approach enables us to deal with the data exclusion problems (cf. Section~\ref{sec:adaptive_detector}) which usually occur due to poor detection performance. 

The adaptive ensemble-based error detection module implements a number of \textit{base error detectors}, such as missing values (MV) detector, outliers detector, duplicates detector, rule violations detector, and ML-based error detectors (cf. Section~\ref{sec:evaluation} for more details). Each base detector generates a list of indices corresponding to the erroneous instances detected by each method. To combine these detections, we implement an \textit{adaptive voting mechanism} where the instances detected by at least $k$ methods are annotated as erroneous. After generating a list of erroneous data instances, a \textit{data sampler} is used to extract a clean data fraction. Before augmenting the clean data fraction, a \textit{data checker} scrutinizes it to detect possible occurrence of the relevant data exclusion problems. If one of these problems is detected, the data sampler iteratively adjusts the value of the voting threshold $k$ so that all relevant data exclusion problems are resolved. 

The extracted clean data fraction is then used as an input to a \textit{variational autoencoder} (VAE) to generate a new set of data from the same distribution. As Figure~\ref{fig:architecture} illustrates, the VAE module implements two feed-forward neural networks, namely an encoder and a decoder. The encoder learns the distribution of the latent space representation of the clean data fraction. Afterward, the decoder uses the sampled latent vector to generate data instances similar to the inputs. In this context, the optimization problem is to minimize (1) the reconstruction loss function which compares the inputs with the decoder-generated values, and (2) the KL divergence which statistically differentiates between the probability distributions of the inputs and the generated data. After generating additional clean data, a \textit{data integrator} merges the newly generated data with the original dirty data. The merged data set is then transformed using feature engineering and wrangling tools, e.g., normalization, embedding, and feature crossing, before training ML models on the transformed data. 
%
In our paper, we assume that the dirty data sets have heterogeneous error profiles. In other words, different error types may simultaneously exist in a data set, e.g., outliers and missing values. To enable clean data augmentation, we also assume that each dirty data set includes a set of clean instances and a set of dirty instances. In the next section, we present the implementation details of the adaptive ensemble-based error detection module.

%% file: sections/error_detection.tex
\section{Adaptive Error Detection}
\label{sec:adaptive_detector}

Before delving into the details of our adaptive ensemble-based error detection method, it is necessary to highlight that the existing \textit{error-dependent} detectors tackles only a subset of the errors in a data set. For instance, a missing value detector finds only null values while overlooking other errors. Similarly, outlier detectors usually employ statistical measures to differentiate between legitimate data and outliers while ignoring other error types, such as rule violation and duplicates. Accordingly, the low detection recall of such detectors prevents them from being individually employed. To maximize the detection recall, there exist several advanced methods which can be subsumed under two major classes, namely the \textit{ML-based methods} and the \textit{ensemble methods}. The first class embodies the detectors which implement semi-supervised binary classifiers to differentiate between clean and dirty data instances \cite{ed219,holodetect19}. The second class comprises the detectors which combine the detections of several error-dependent methods \cite{max_min16} to improve the detection recall together with providing a knob for controlling the detection precision (defined as the fraction of relevant instances, e.g., actual dirty data instances, among the detected instances). 

\begin{figure}[tbph]
	\centering
	\subfloat[Attribute-level]{\label{fig:attribute}\includegraphics[width=0.35\columnwidth]{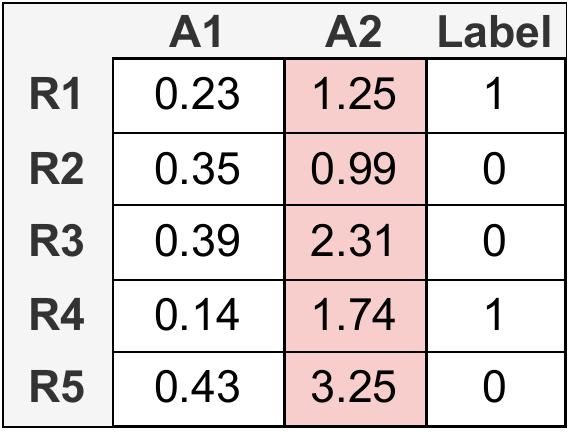}} \hspace{3mm}
	\subfloat[Class-level]{\label{fig:class}\includegraphics[width=0.35\columnwidth]{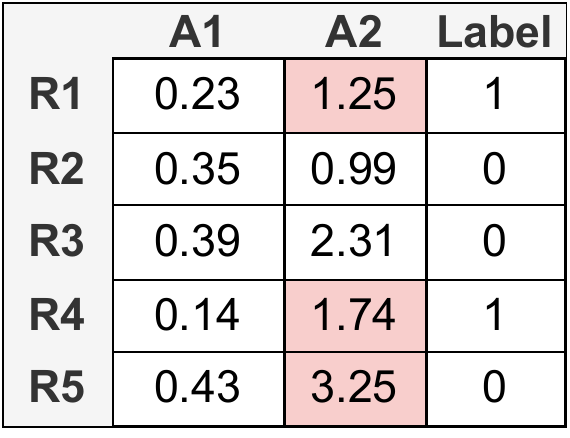}} \hfill
	\caption{Examples of data exclusion problems}
	\label{fig:exclusion}
\end{figure}

Nevertheless, ML-based and ensemble detectors usually suffer from \textit{relevant data exclusion} problems when the main objective is to extract a clean fraction from the dirty data set, as required in \PaperAcronym. Specifically, data exclusion problems typically arise when the extracted clean fraction is either (1) empty or (2) does not comprise all classes (i.e., unique labels in a data set) which exist in the dirty data set. The first type, referred to as the \textit{attribute-level} exclusion, commonly emerges when an error detector flags all data instances of a certain attribute as erroneous (cf. red instances in the attribute $A2$ in Figure~\ref{fig:attribute}). In this case, the data sampler will fail to extract a clean fraction because all records contain erroneous data instances. The second type, referred to as the \textit{class-level} exclusion, occurs when an error detector flags all data instances from one or more classes as erroneous (cf. red instances in Figure~\ref{fig:class}). Accordingly, these classes (e.g., class \quotes{1} in Figure~\ref{fig:class}) will not be represented in the clean fraction, which in turn broadly changes the data distribution of the augmented data. In particular, both data exclusion problems may arise due to (1) the false positives of some error-dependent or ML-based detectors, and (2) fixing the value of the voting threshold $k$, in ensemble detectors, for all data instances. 

\begin{figure}[tbph]
	\centering
	\includegraphics[width=\linewidth]{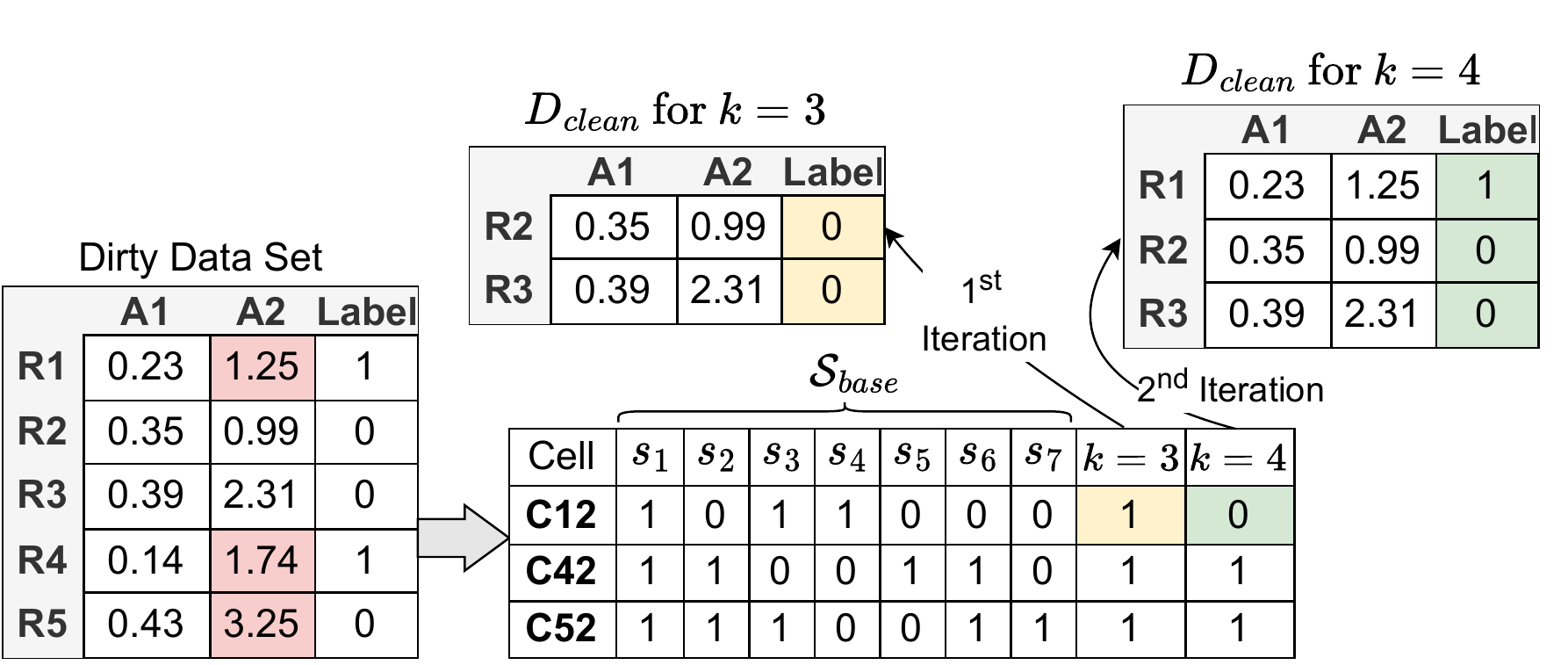}
	\caption{Example of adapting the voting threshold}
	\label{fig:voting_example}
\end{figure}

To overcome the data exclusion problems, \PaperAcronym introduces an \textit{adaptive} ensemble method which can dynamically adjust the voting threshold to preserve all classes in the extracted clean fraction. To this end, the proposed adaptive ensemble-based method makes the detection decisions iteratively. Specifically, the data sampler adjusts the value of the voting threshold $k$ whenever it detects the occurrence of data exclusion problems. Figure~\ref{fig:voting_example} shows an example of tuning the voting threshold $k$ while detecting errors in a dirty toy data set, which consists of five records $R1$-$R5$ and two attributes $A1$ and $A2$. After running all available detectors, i.e., $S1$-$S7$, the iterative voting mechanism begins with an initial value of the voting threshold, i.e., $k=3$. In this scenario, the cell $C12$ has been detected by three detection methods, i.e., $s_1$, $s_3$ and $s_4$, while the cell $C42$ has been detected by four detection methods, i.e., $s_1$, $s_2$, $s_5$ and $s_6$.

According to the traditional Min-K voting mechanism, the cells $C12$ and $C42$ are flagged as erroneous since they have been detected by at least three methods. Consequently, the records containing these cells are to be excluded from the clean fraction, as shown on the top left table (i.e., $D_{clean}$ for $k=3$).  As a result, a class-level data exclusion problem emerges, where the clean fraction lacks records with the class \quotes{1}. To resolve this problem, \PaperAcronym updates the voting threshold $k$ from three to four in the second iteration. In this case, the cell $C12$ does not exceed the threshold. Accordingly, it is flagged as a clean cell and thus its record will be included in the clean fraction, as shown on the top right table (i.e., $D_{clean}$ for $k=4$). 

To clarify the implementation details of \PaperAcronym, Figure~\ref{fig:alg_detector} demonstrates the pseudocode of the proposed data curation pipeline. At the outset, \PaperAcronym implements an inventory of several error-dependent and ML-based detectors $\mathcal{S}_{base}$. Each detector $s_{i}$ generates a list of indices of the erroneous instances, e.g., $R_{i} = \{(r_u,a_v)\}$, where $r_u$, $a_{v}$ are the $u$th record and the $v$th attribute in the dirty data set $D_{dirty}$. Afterward, all detections of $m$ detectors $R_{all}$ are merged, before estimating the cells count $Q_{cells}$, i.e., which quantifies for each cell $c_j$, the number of times they have been detected (cf. lines~\ref{line:count_1}-\ref{line:count_2}). Initially, \PaperAcronym assigns the current voting threshold $k_{cur}$ a small value which is larger than one (e.g., $k_{cur}=2$ while setting $k_{cur}=1$ implies using only one error-dependent or ML-based detection method). In the first iteration, the algorithm uses an initial value, i.e., $k_{cur} = k_{init}$, to decide for each cell whether it is erroneous (cf. lines \ref{line:cur_1} and \ref{line:cur_2}). The final list of detections $R_{ensemble}$ is then utilized to extract a clean fraction $D_{clean}$, as shown in line~\ref{line:extract}. 
\begin{figure}	
	\begin{algorithmic}[1]
		\Require dirty data set $D_{dirty}$, initial voting threshold $k_{init}$, augmentation budget $N_{aug}$, base detectors $\mathcal{S}_{base} = s_1, \cdots, s_m$, update rate $\alpha_u$
		
		\ForAll{detector $s_i \in \mathcal{S}_{base}$} \Comment{Running all available detectors}
		\State \textbf{Estimate} detections $R_{i} \gets s_i(D_{dirty})$
		\State \textbf{Append} $R_{i}$ to the list of all detections $R_{all}$  
		\EndFor
		\State \textbf{Initialize} the cells counter $Q_{cells} \gets 0$
		\ForAll{cell $c_{j} \in R_{all}$}  \Comment{Estimating the cells count} \label{line:count_1}
		\If{cell $c_{j}$ in cells counter $Q_{cells}$}
		\State \textbf{Increment} the counter $Q_{cells}[c_{j}] \gets Q_{cells}[c_{j}] + 1$ \label{line:count_2}
		\EndIf
		\EndFor
		\State \textbf{Initialize} the list of missing classes $L_{miss} \gets \emptyset$
		\State \textbf{Initialize} the temporary thresholds $k_{attr}, k_{class}\gets k_{init}$
		\While{true}
		\ForAll{cell $c_{j} \in Q_{cells}$}	\Comment{Adaptive Min-K voting}
		
		\If{\textit{classes}($c_{j}) \in L_{miss}$} \label{line:check}
		\State \textbf{Assign} current threshold $k_{cur} \gets k_{class}$
		\Else \State \textbf{Assign} current threshold $k_{cur} \gets k_{attr}$
		\EndIf
		
		\If{$Q_{cells}[c_{j}] \geq k_{cur}$} \label{line:cur_1}
		\State \textbf{Append} cell $c_{j}$ to the final detection list $R_{ensemble}$ \label{line:cur_2}
		\EndIf
		\EndFor
		\State \textbf{Extract} clean fraction $D_{clean} \gets D_{dirty} \backslash R_{ensemble}$ \label{line:extract}
		
		\If{$D_{clean} = \emptyset$} \Comment{Attribute-level data exclusion} \label{line:exclusion_1}
		\State \textbf{Update} threshold $k_{attr} = k_{attr} + \alpha_u$ \label{line:attr}
		\ElsIf{\textit{classes}($D_{clean}$) $\neq$ \textit{classes}($D_{dirty}$)}	\Comment{Class-level}
		\State \textbf{Find} $L_{miss} \gets$ \textit{classes}($D_{dirty}$) $\backslash$ \textit{classes}($D_{clean}$) \label{line:class_1}
		\State \textbf{Update} threshold $k_{class} = k_{class} + \alpha_u$ \label{line:class_2}
		\If{the difference $|K_{class} - k_{attr}| \geq 2\times\alpha_u$}
		\State \textbf{Update} threshold $k_{attr} = k_{attr} + \alpha_u$
		\EndIf 
		\Else \Comment{No data exclusion problems exist} 
		\State \textbf{Break} \label{line:exclusion_2}
		\EndIf
		\EndWhile
		\State \textbf{Generate} additional data $D_{aug} \gets$ \textit{run\_vae}($D_{clean}, N_{aug}$)
		\State \textbf{Combine} data sets {$D_{final} \gets D_{dirty} \cup D_{aug}$}
	\end{algorithmic}
	\caption{Ensemble-based error detection algorithm 
		\label{fig:alg_detector}}	\vspace{-4mm}
\end{figure}

Before running the variational autoencoder, the algorithm checks for the occurrence of data exclusion problems (cf. lines~\ref{line:exclusion_1}-\ref{line:exclusion_2}). An attribute-level data problem can be detected if the size of the clean data fraction is equal to zero. If an attribute-level problem is detected, then the algorithm increases the temporary threshold $K_{attr}$ by the value of the update rate $\alpha_u$ (cf. line~\ref{line:attr}). Increasing the threshold clearly implies relaxing the voting mechanism, which in turn reduces the number of detected cells. In the subsequent iteration, the algorithm re-estimates the list of all detections $R_{all}$ and the clean fraction $D_{clean}$. Next, it checks again for possible attribute-level data exclusion, before inspecting $D_{clean}$ for class-level exclusion problems. To this end, it simply compares the number of classes in the dirty data \textit{classes}($D_{dirty}$) and in the clean fraction \textit{classes}($D_{clean}$). If a class-level problem is detected, \PaperAcronym estimates the list of missing classes $L_{miss}$ (cf. lines~\ref{line:class_1}-\ref{line:class_2}). In this case, \PaperAcronym deliberately differentiates between the cells belong to records whose classes are included in $L_{miss}$, e.g., $C12$ and $C42$ in Figure~\ref{fig:voting_example}, and all other cells, e.g., $C52$. 

The rationale behind this differentiation is to achieve a high detection precision while ensuring that all classes in $D_{dirty}$ also exist in the clean fraction $D_{clean}$. To this end, \PaperAcronym relaxes the voting algorithm for the cells belonging to the missing classes, e.g., $C12$ and $C42$, to facilitate their inclusion in the clean fraction. Whereas, \PaperAcronym becomes more strict with other cells, e.g., $C52$, since their classes are already represented in the clean fraction. For this purpose, \PaperAcronym adopts two different temporary thresholds $k_{attr}$ and $k_{class}$. It updates the threshold $k_{class}$ when a class-level exclusion occurs. In the next iteration, the algorithm checks whether each cell $c_j$ belongs to the records of the missing classes (cf. line~\ref{line:check}). If that is the case, it employs the temporary threshold $k_{class}$, which is always greater than $k_{attr}$ by the value of the update rate $\alpha_u$, for making the detection decisions. Otherwise, it uses the threshold $k_{attr}$. The algorithm keeps iterating till all data exclusion problems are resolved. Afterward, the final clean fraction is merged with the dirty data, before carrying out the typical wrangling and feature engineering tasks.

%% file: sections/augmentation.tex
\section{Clean Data Augmentation}
\label{sec:augmentation}

In this section, we discuss the technical details of the clean data augmentation module. After extracting a clean data fraction, a data augmentation module is used to generate data from the same distribution as the clean fraction. In this regard, we examined three data augmentation methods dedicated to tabular data, including MODALS \cite{cheung2020modals}, Variational Autoencoders (VAE) \cite{kingma2013auto}, and Conditional Tabular Generative Adversarial Network (CTGAN) \cite{goodfellow2020generative}. Through our experiments, we found that VAE outperforms the other two methods\footnote{For brevity, we omitted the results of our comparative study between the three augmentation methods in this paper.}. Specifically, we examined the three methods in an ML pipeline for several datasets. The VAE data augmentation achieved higher predictive accuracy (F1 scores $\geq$ 0.98\%) than the other two methods. Therefore, we decided to integrate the VAE method with our adaptive error detection.

In general, an autoencoder implements two components, namely an encoder and a decoder. The encoder compresses the input data set in a low dimensional space, referred to as the \textit{latent space} representation. Afterward, the decoder exploits the latent space representation to recover the input data set. For this purpose, autoencoders typically have a reconstruction loss function to compare the original data with the values generated by the decoder. For data augmentation, autoencoders perform some variations in the latent space. To this end, the VAE module trains its encoder to extract the parameters of data distributions rather than the latent space representation \cite{kingma2013auto}. As depicted in Figure~\ref{fig:architecture}, the decoder generates a two-dimensional vector consisting of the mean $\mu_x$ and the variance $\sigma_x$. Afterward, VAE generates a set of data instances from the Gaussian distribution $\mathcal{N}(\mu_{x},\sigma_{x})$ formed using the extracted parameters. Such a set of data instances are then used as an input to the decoder to generate additional clean data. 

In addition to the reconstruction loss, the VAE module employs the KL divergence to distinguish between the probability distributions of the original data set (i.e., the clean fraction in \PaperAcronym) and the generated data. Generally, the KL divergence is a statistical measure which quantifies how a probability distribution differs from a reference distribution. The VAE module strives to reduce the value of KL divergence via optimizing the mean and variance to simulate the input distribution.  
Aside from the optimization metrics, the process of sampling data from a distribution, parameterized by the generated means and variances, is not differentiable. In this case, it is relatively challenging to perform backpropagation (necessary to optimize the weights of the encoder and the decoder) over the random node $Z$. To overcome this problem, the reparameterization process is performed to enable backpropagation through the random node. To this end, the reparameterizer turns the random node $Z \sim \mathcal{N}(\mu_x,\sigma_x)$  into a differentiable function $Z = \mu + \sigma \odot \epsilon$, where $\epsilon \sim \mathcal{N}(0,1)$ represents the standard Gaussian distribution, and it is irrelevant for taking the gradients (which is a necessary step in the backpropagation process).

%% file: sections/evaluation.tex
\section{Performance Evaluation}
\label{sec:evaluation}
In this section, we assess the effectiveness of \PaperAcronym relative to a set of baseline methods. Through the experiments, we seek to answer the following questions: (1) How does \PaperAcronym compare to the baseline methods in terms of their impact on downstream ML models (the performance of such models is quantified in terms of the predictive accuracy and the training time)? (2) What is the minimum amount of augmented clean data required to yield a comparable performance as the ground truth? (3) What is the impact of the voting threshold $k$ on the detection accuracy? (4) what is the impact of the error rate of a data set on the performance of \PaperAcronym? (5) How does \PaperAcronym contribute to the model performance? We first describe the setup of our evaluations, before discussing the results and the lessons learned throughout this study.
\subsection{Experimental Setup}
%
We evaluated \PaperAcronym using six real-world data sets covering different data sizes and different error rates. Table~\ref{tab:datasets} summarizes the characteristics of each data set. Three of such data sets, i.e., Adult, breast Cancer, and Smart Factory, are associated with classification (CL) tasks. Whereas, the other three data sets, i.e., Nasa, Housing, and Soil Moisture, have regression tasks (REG). To control the evaluation environment, we opted to inject different realistic errors into the data sets. To this end, \PaperAcronym leverages the BART tool \cite{bart15} which provides a systematic control over the amount of errors and how hard these errors are to be repaired. To inject errors using BART, we use a set of denial constraints to generate different error types, such as rule violation, outliers, and missing values. 
\begin{table}[tbph]
	\centering
	\caption{Data sets used in the evaluation where $\gamma$ is the error rate (\%) and the error types are rules violation (RV), outliers (OT), missing values (MV), and typos (TP)}
	\label{tab:datasets}
	\resizebox{\columnwidth}{!}{%
		\begin{tabular}{lccccc}
			\toprule
			\multicolumn{1}{l}{\textbf{Data Set}} &
			\multicolumn{1}{l}{\textbf{Rows}} &
			\multicolumn{1}{l}{\textbf{Columns}} &
			\multicolumn{1}{l}{\textbf{Error Types}} &
			\multicolumn{1}{l}{\textbf{Error Rate}} &
			\multicolumn{1}{l}{\textbf{ML Task}} \\ \midrule
			Adult \cite{adult96}                & 45223 & 15  & RV, OT     & 0.09  & CL   \\
			Breast Cancer \cite{ucidata19}      & 700   & 12  & MV, TP, OT & 0.4  & CL   \\
			Smart Factory \cite{smartfactory22} & 23645 & 19  & MV, OT     & 0.83 & CL   \\
			Nasa \cite{nasa22}                  & 1504  & 6   & MV, OT, TP & 0.13  & REG  \\  
			Housing \cite{housing93}                 & 507 & 14  & MV, OT     & 0.24   & REG  \\ 
			Soil Moisture \cite{soilmoisture18} & 679   & 129 & MV, OT     & 0.3  & REG  \\ \bottomrule   
		\end{tabular}
	}
\end{table}

We compare the performance of \PaperAcronym to a wide collection of baseline methods composed of different error detection and repair tools. Table~\ref{tab:baselines} lists the error detection and repair methods. The combination of each detection and repair method represents a baseline method. For instance, the baseline B2 implies detecting errors in a dirty data set using dBoost and generating repair candidates using an ML-based imputation algorithm. In the list of repair methods, we also include the ground truth and dirty cases to show the performance upper- and lower-bound. The standard imputer replaces the erroneous data instances with the mean values of the numerical columns and with a dummy string for the categorical columns. Alternatively, the ML-based imputer employs a decision tree model to predict the erroneous numerical instances, while it employs missForest \cite{missforest12} for categorical data. The evaluation metrics used in the evaluation comprise the detection accuracy, i.e., expressed in terms of the detection precision, recall, and F1 score, the training time, and the predictive accuracy.
\begin{table}[tbph]
	\centering
	\ra{1.1}
	\caption{Error detection and repair tools used as baselines}
	\label{tab:baselines}
	\resizebox{\columnwidth}{!}{%
		\begin{tabular}{@{}cccccc@{}}
			\toprule
			\multicolumn{1}{l}{\textbf{Idx}} &
			\multicolumn{1}{l}{\textbf{Detectors}} &
			\multicolumn{1}{l}{\textbf{Idx}} &
			\multicolumn{1}{l}{\textbf{Detectors}} &
			\multicolumn{1}{l}{\textbf{Idx}} &
			\multicolumn{1}{l}{\textbf{Repair Method}} \\
			\midrule
			B & dBoost \cite{dboost16}      & M & Min-K \cite{max_min16}        & - & Dirty            \\
			E & ED2 \cite{ed219}          & N & NADEEF \cite{nadeef13}       & 1 & Ground Truth     \\
			F & FAHES \cite{fahes18}       & Q & Outliers: IQR & 2 & ML Imputer       \\
			H & HoloClean \cite{holoclean17}   & R & RAHA  \cite{raha19}        & 3 & Standard Imputer \\
			I & Outliers: IF \cite{outliers13} & S & Outliers: SD & 4 & BARAN \cite{baran20}           \\
			K & KATARA \cite{katara15}      & V & MV-Detector   &   &      \\ \bottomrule            
		\end{tabular}%
	}
\end{table}

For the VAE module, both the encoder and the decoder have been implemented as feed-forward neural networks. The inputs to the VAE module are a training set, a test set, the dimensions of the clean fraction, the number of nodes in the hidden layers, and the number of latent factors. The number of nodes in the input layer of the encoder and in the output layer of the decoder has been set to the number of attributes in the clean fraction. The encoder and decoder comprise two hidden layers, and the number of nodes in the first and second hidden layers has been set to 50 and 12, respectively. We employ the Adam optimizer to optimize the parameters and custom loss, which are the combined mean square error and the KL divergence. We add the ReLU activation function for each hidden layer in the encoder and the decoder.

To examine the performance of \PaperAcronym and the baseline methods, we implemented a neural network, using Keras, which can deal with regression, binary, and multi-class classification tasks\footnote{The source code of \PaperAcronym, the baseline methods, and the data sets, are to be publicly released with the final version of the paper.}. \PaperAcronym leverages a Bayesian-based informed search method, referred to as Optuna \cite{optuna2019}, to tune the learning rate, the number of hidden layers, and the number of units per layer. For all experiments, we fixed the number of training epochs to 500.
All experiments have been repeated ten times, where the means of the ten runs are reported. For clarity, we omitted the standard deviations when they are less than 1\%. 
We run all the experiments on an Ubuntu 16.04 LTS machine with 32 2.60 GHz cores and 128 GB memory. 

\subsection{Results}
\begin{figure}[tbph]
	\centering
	\subfloat[Housing]{\label{fig:housing_threshold}\includegraphics[width=0.5\columnwidth]{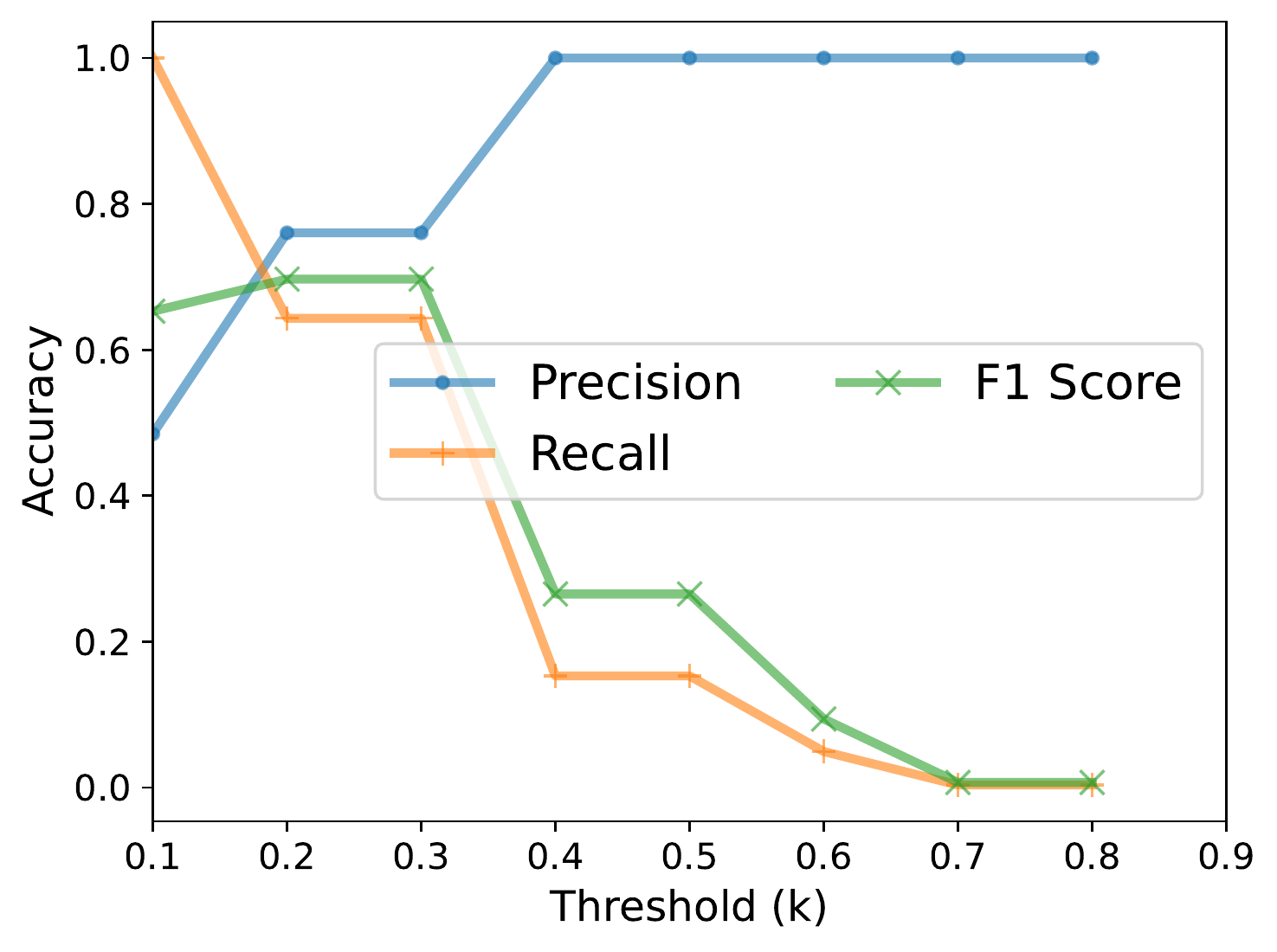}} \hfill
	\subfloat[Smart Factory]{\label{fig:factory_threshold}\includegraphics[width=0.5\columnwidth]{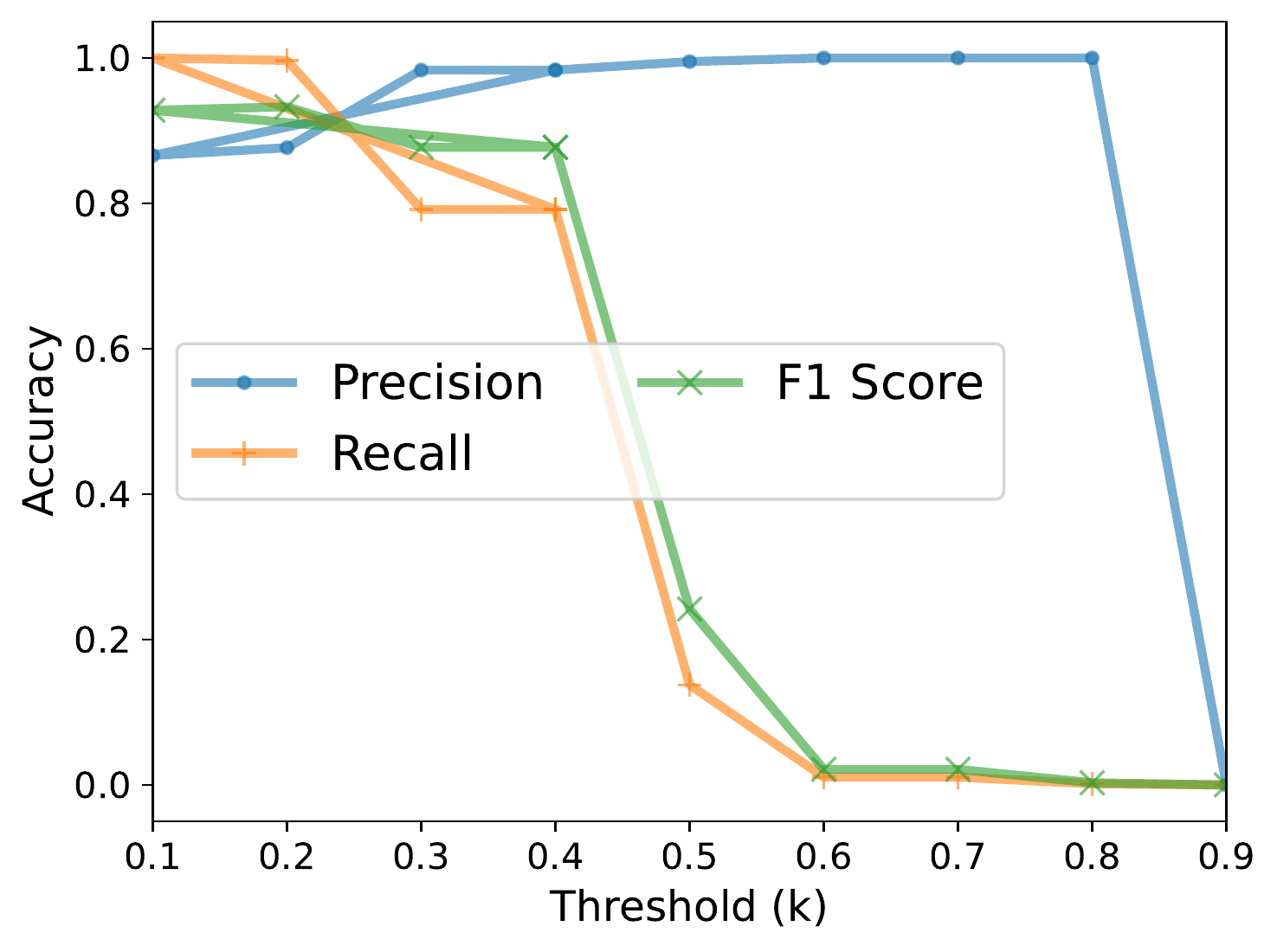}} \hfill
	\caption{Detection accuracy against different values of the threshold $k$ (Min-k)}
	\label{fig:min_k_threshold}
\end{figure}

%
\paragraph{Voting Threshold}
%
Before delving into the results obtained for \PaperAcronym and the baseline methods, we first assess the impact of changing the voting threshold to properly motive for the adaptive voting mechanism implemented in \PaperAcronym. Figures~\ref{fig:housing_threshold} and \ref{fig:factory_threshold} depict the detection accuracy of the traditional Min-k detection method while detecting erroneous instances in the Housing and Smart Factory data sets, respectively. Both figures show that increasing the voting threshold $k$ has distinct influences on the detection recall (defined in Section~\ref{sec:concept}) and the detection precision (defined in Section~\ref{sec:adaptive_detector}). For instance, Figure~\ref{fig:housing_threshold} demonstrates that increasing the voting threshold $k$ usually leads to reducing the detection recall due to annotating fewer data instances as erroneous. On the other hand, increasing the threshold $k$ causes the precision to be improved thanks to the precise estimation of erroneous instances using a consensus among multiple error detection methods. In light of this trade-off, it is not often straightforward to accurately select the optimal value of $k$ for a certain application scenario. Moreover, we noticed during our experiments that hard-coding the value of $k$, as implemented in the traditional Min-K method, typically prevents us from extracting a \textit{balanced} clean data fraction. Therefore, \PaperAcronym iteratively adjusts the value of $k$ to combat the relevant data exclusion problems. 

\begin{figure*}[bhtp]
	\centering
	\subfloat[Adult (F1 Score)]{\label{fig:adult_robust}\includegraphics[width=0.33\linewidth]{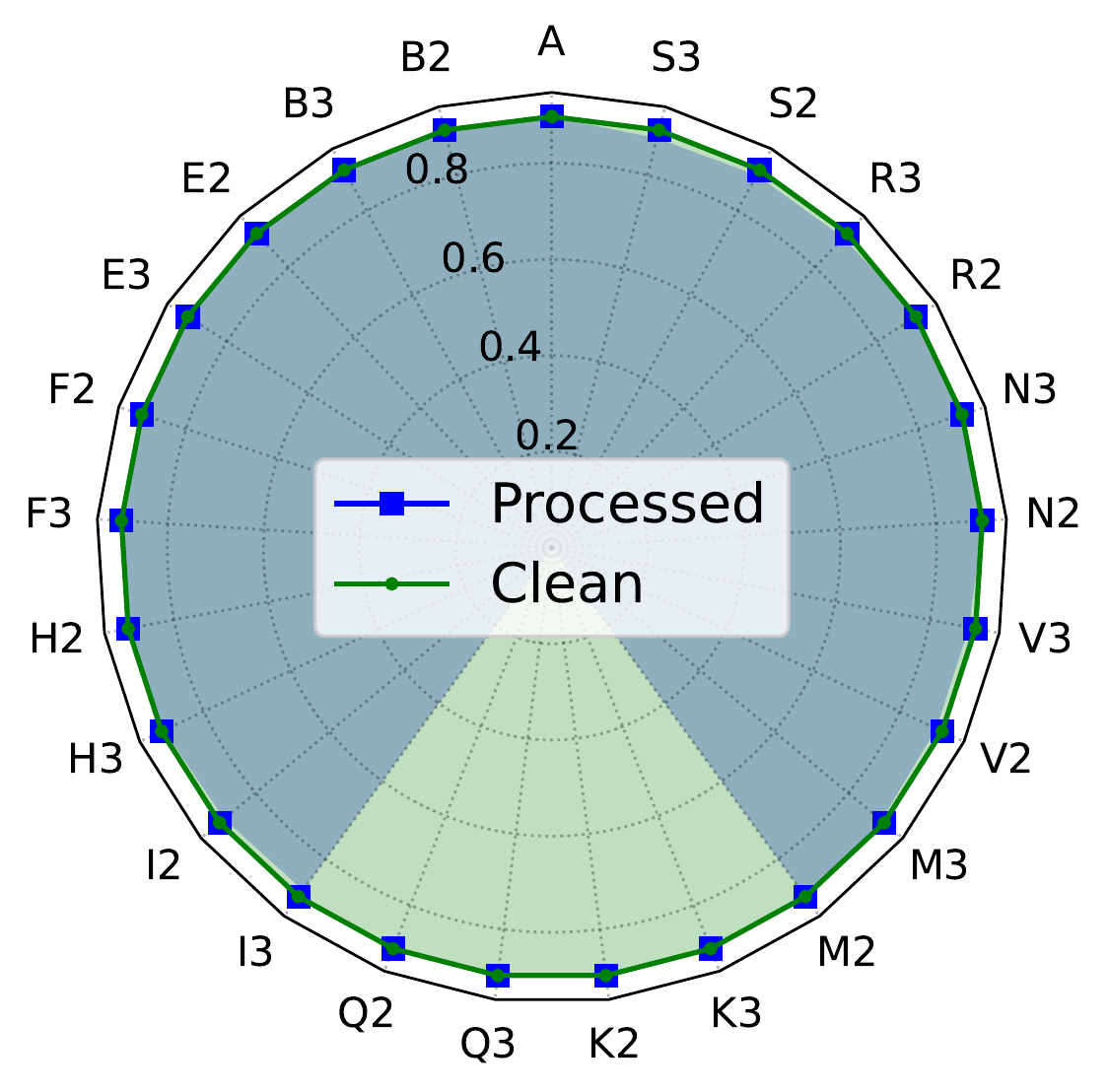}} \hfill
	\subfloat[Breast Cancer (F1 Score)]{\label{fig:bcancer}\includegraphics[width=0.33\linewidth]{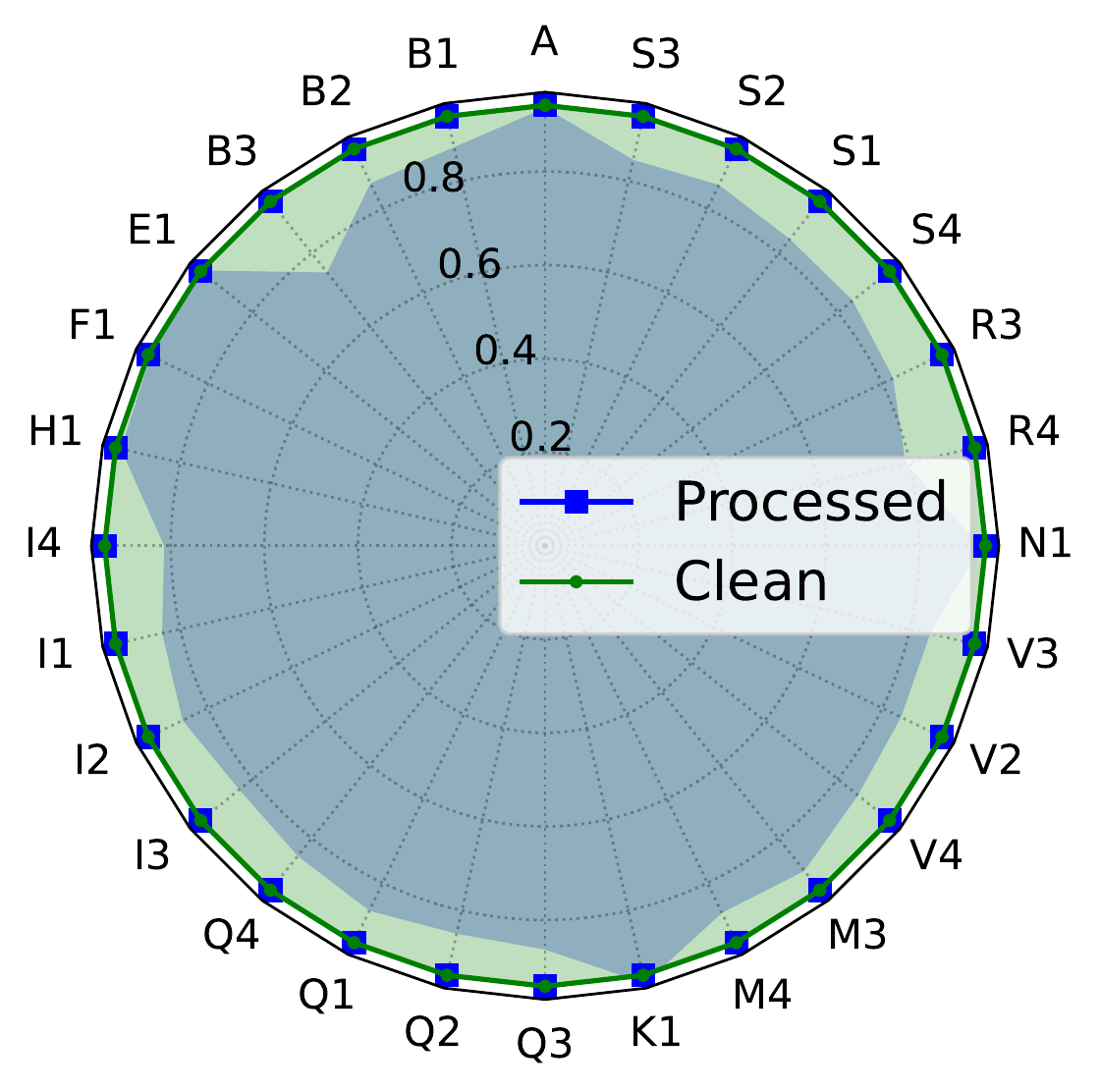}} \hfill
	\subfloat[Smart Factory (F1 Score)]{\label{fig:factory}\includegraphics[width=0.33\linewidth]{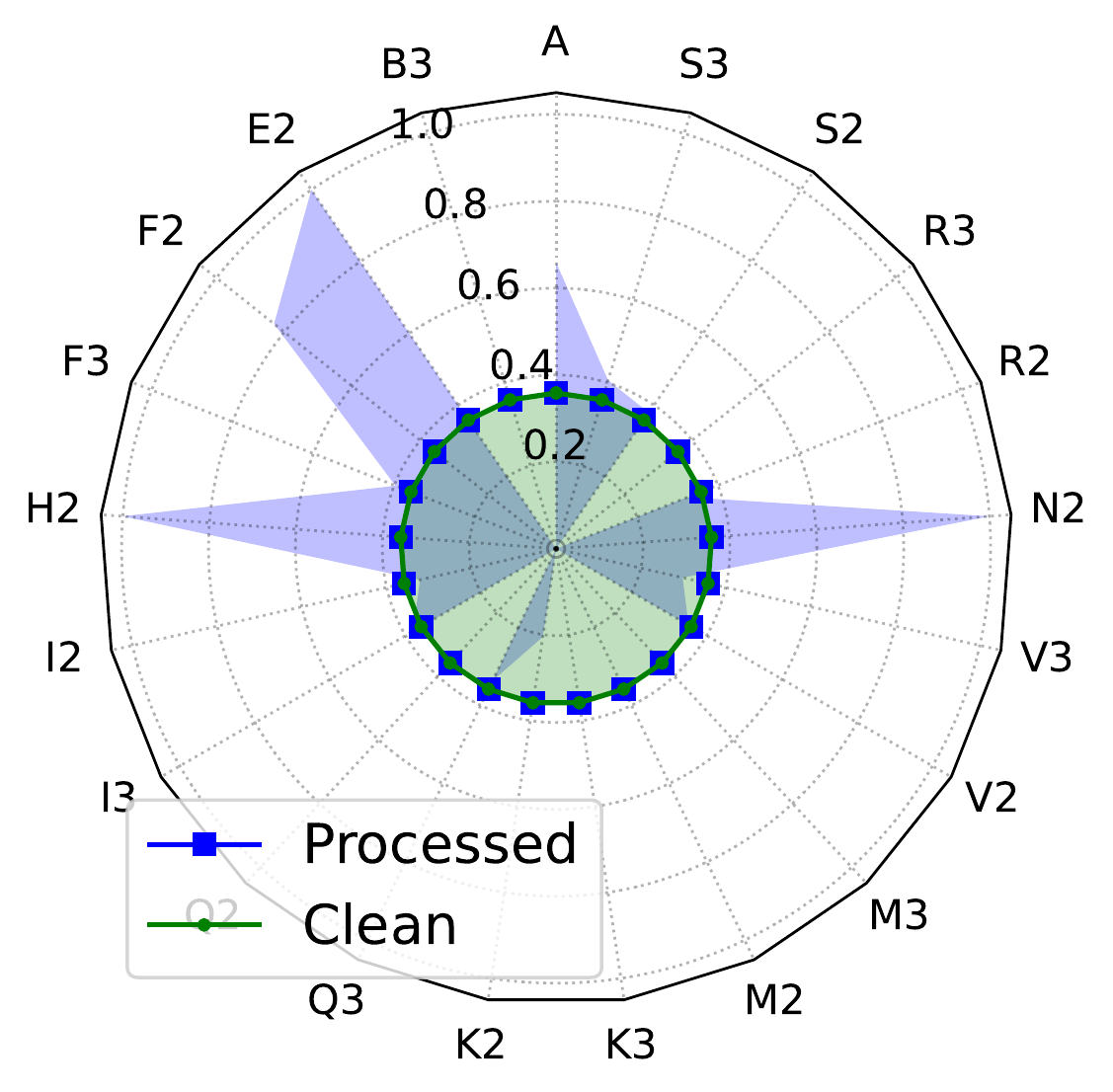}} \hfill
	\subfloat[Nasa (MSE)]{\label{fig:nasa}\includegraphics[width=0.33\linewidth]{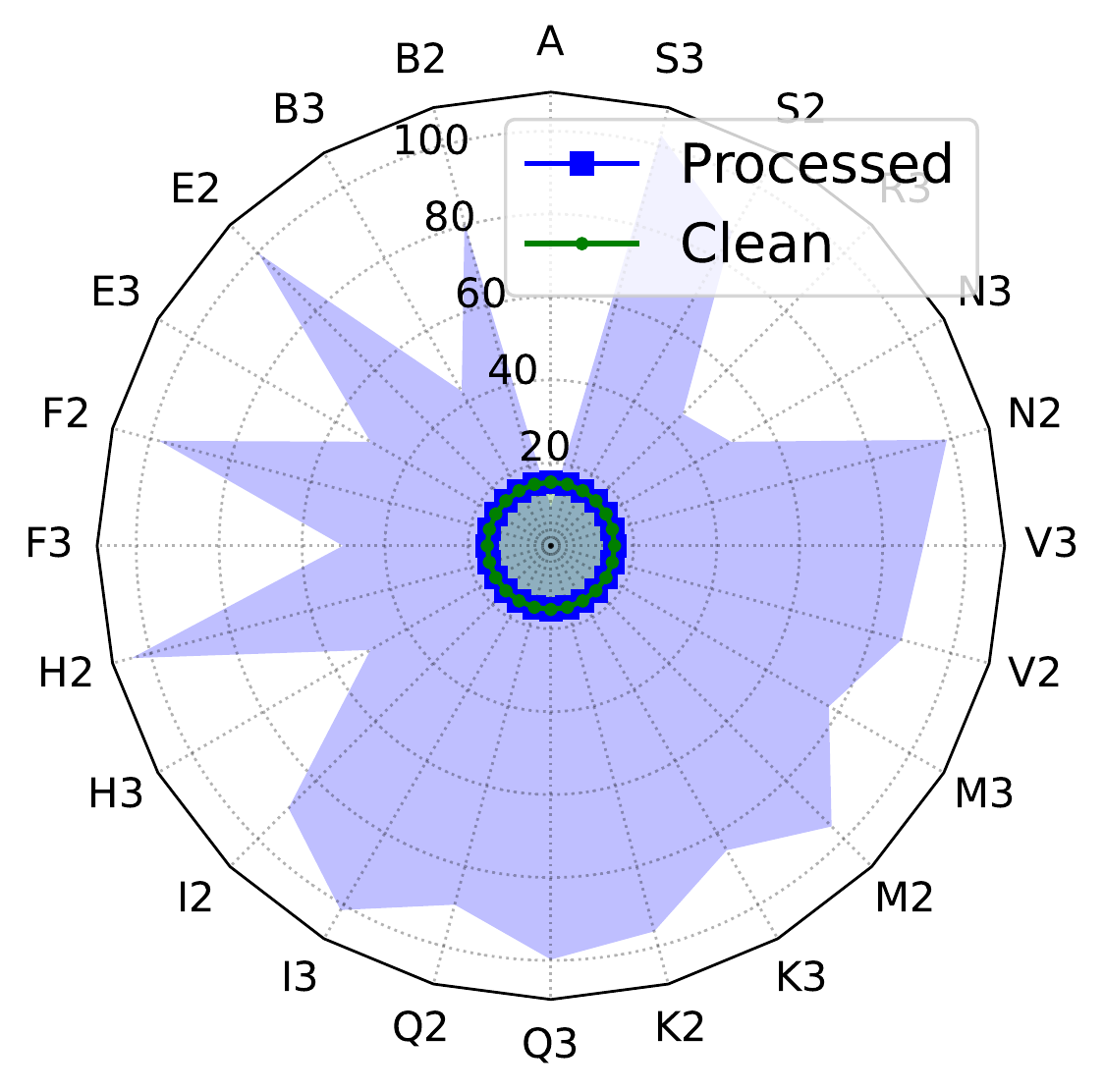}} \hfill
	\subfloat[Soil Moisture (MSE)]{\label{fig:moisture}\includegraphics[width=0.33\linewidth]{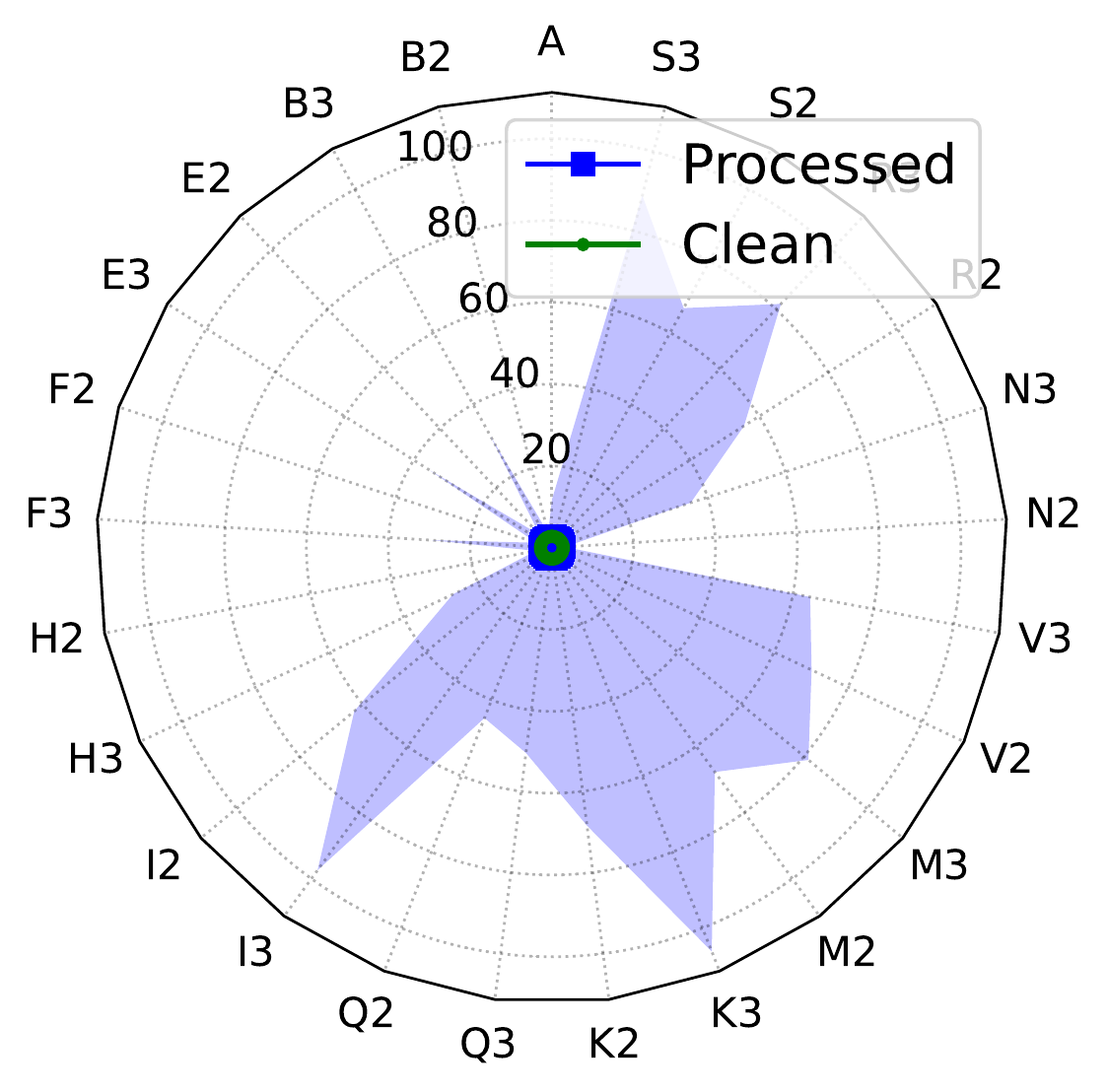}} \hfill
	\subfloat[Housing (MSE)]{\label{fig:housing}\includegraphics[width=0.33\linewidth]{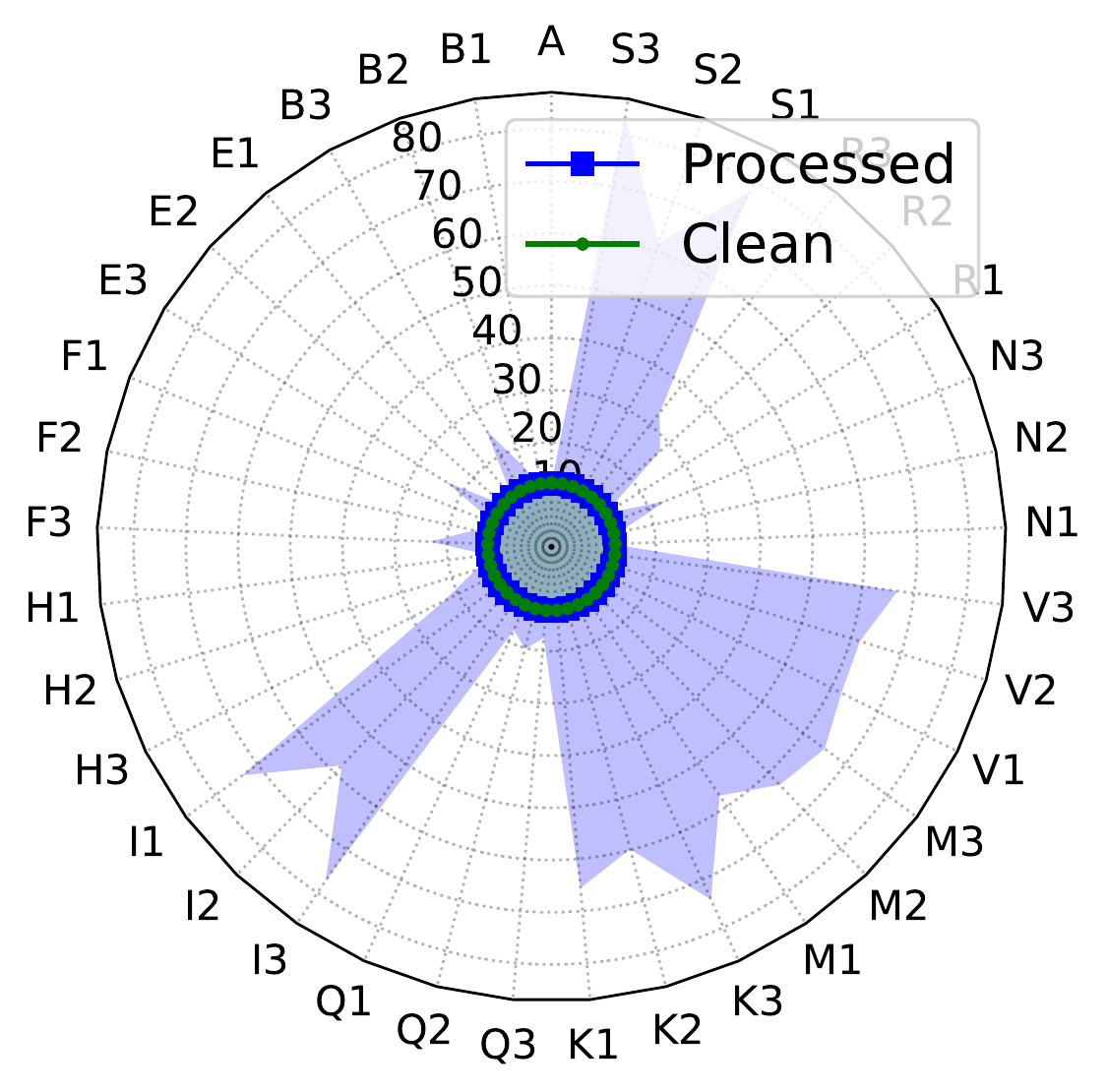}} \hfill 
	\caption{Modeling accuracy of clean datasets, \PaperAcronym (denoted as \quotes{A}), and a set of baseline methods}
	\label{fig:radar_figs}
\end{figure*}

\paragraph{Predictive Accuracy} In this set of experiments, we estimate the accuracy of a neural network trained on different versions of the examined data sets. For \PaperAcronym, we set the number of generated instances to 6000 for all data sets. Figure~\ref{fig:radar_figs} depicts the predictive performance of a neural network trained on a clean data set\footnote{In this context, a clean data set means the ground truth version of the data.} (cf. green area in Figure~\ref{fig:adult_robust}) and a set of processed data sets (cf. blue area in Figure~\ref{fig:adult_robust}). The processed data sets comprise the data sets curated by \PaperAcronym (abbreviated as A in Figure~\ref{fig:radar_figs}) and the ones curated using different combinations of baseline tools. For the Adult data set, Figure~\ref{fig:adult_robust} compares the performance of \PaperAcronym, the clean data set, and 24 baseline methods in terms of the F1 score of the neural network. As the figure depicts, \PaperAcronym achieves similar accuracy as the clean data set (both achieve an average F1 score of 89.5\%). It is important to notice that some curation combinations yield a similar performance as \PaperAcronym, while others, e.g., Q2, Q3, K2, and K3, fail to accurately curate the dirty data set. 

Figure~\ref{fig:bcancer} illustrates a similar evaluation of \PaperAcronym and the baseline methods using the Breast Cancer data set. Again, \PaperAcronym outperforms all baseline methods and achieves a similar performance as the clean data set (an average F1 score of 93.6\% in case of \PaperAcronym and 94\% in case of the clean data set). Figure~\ref{fig:factory} depicts an evaluation of the compared methods using the Smart Factory data set. Obviously, \PaperAcronym outperforms most baseline methods as well as the clean data set (on average by 46\%). It is important to highlight that some baseline methods, e.g., E2, H2, and N2, achieve an F1 score of one. However, such a performance of these baseline methods broadly depends on the data set.

For the Nasa data set, \PaperAcronym outperforms most baseline methods (an average MSE of 9.3 in case of \PaperAcronym and 15.3 in case of the clean data set). For the Soil Moisture data set, \PaperAcronym achieves a reasonable accuracy (an average MSE of 11.7) relative to the clean version (an average MSE of 2.78) and some baseline methods (average MSE of 3.15 and 3.55 for N2 and H2, respectively). In fact, it is important to highlight the consistency of the results obtained by \PaperAcronym over different data sets. While, the other baseline methods lack such a consistency, e.g., N2 and H2 perform poorly in case of the Nasa data set (Figure~\ref{fig:nasa}). Finally, Figure~\ref{fig:housing} shows the results, in terms of MSE, in case of the Housing data set. Again, \PaperAcronym achieves a similar performance as the clean data set (both achieve an average MSE of 12.2).

\paragraph{Training Time} Aside from the predictive accuracy, we also evaluate \PaperAcronym and the baseline methods in terms of the training time. Table~\ref{tab:time} lists the average training times of the neural network trained on the clean and the curated data sets. Due to data augmentation, \PaperAcronym causes the training time to be relatively increased for all data sets. However, for some data sets, such as Adult, the training time in case of \PaperAcronym is slightly higher than that using the clean data set (on average by circa 2.9 minutes). Moreover, a fair comparison should consider a realistic scenario, where data scientists typically lack knowledge about which curation tools to implement in their ML pipelines. Due to the existence of multiple traditional data curation methods, it becomes necessary to examine and execute multiple traditional data curation methods, as in BoostClean \cite{boostclean17} and ActiveClean \cite{activeclean16}. Accordingly, a practical estimation of the training time, caused by the traditional curation methods, should consider a combined estimation of all training times. In this case, \PaperAcronym significantly outperforms the traditional curation methods (named \quotes{Combined} in Table~\ref{tab:time}) for all data sets.

\begin{figure*}[t]
	\centering
	\subfloat[Adult]{\label{fig:adult_aug}\includegraphics[width=0.2\linewidth]{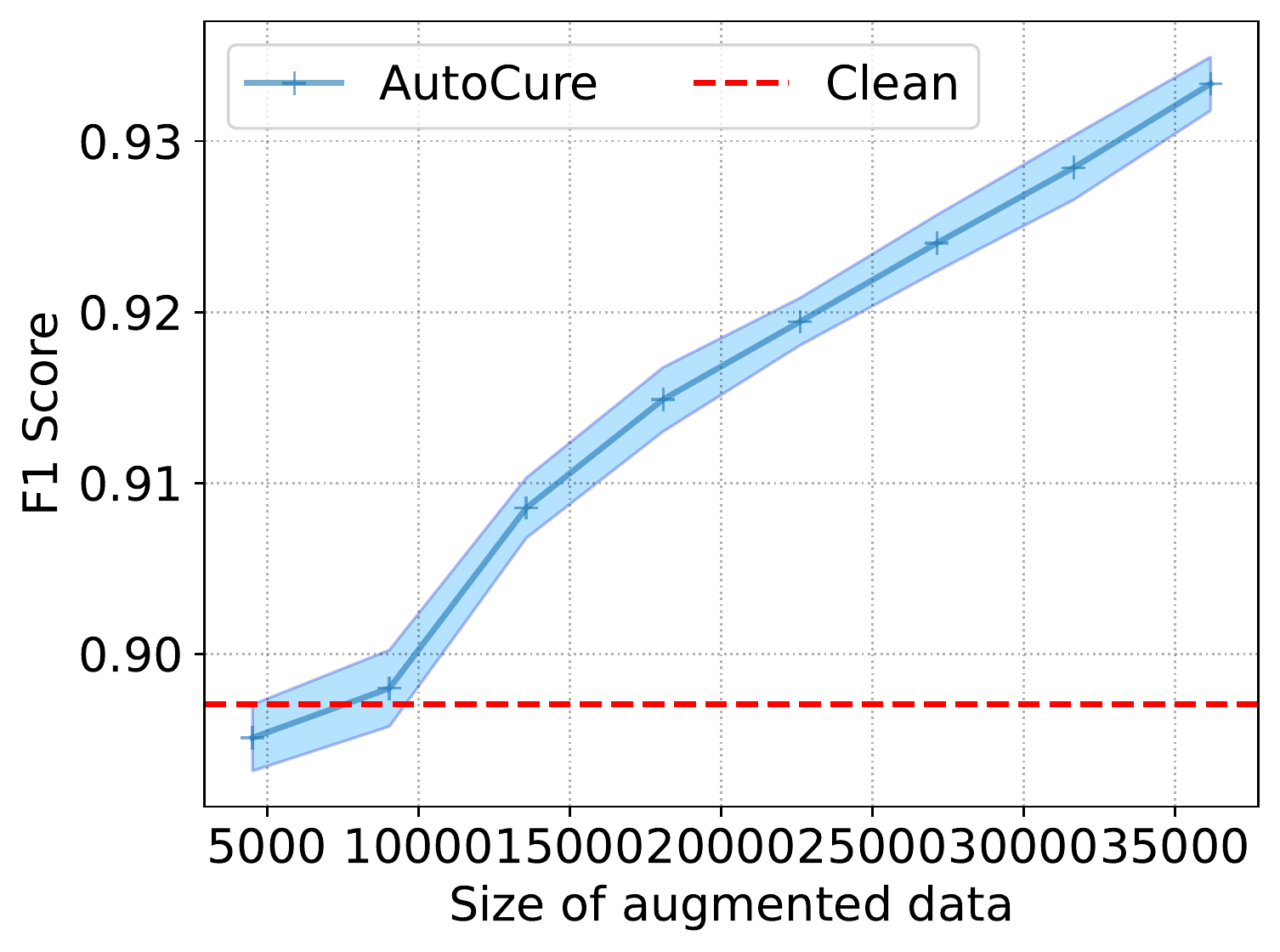}} \hfill
	\subfloat[Smart Factory]{\label{fig:factory_aug}\includegraphics[width=0.2\linewidth]{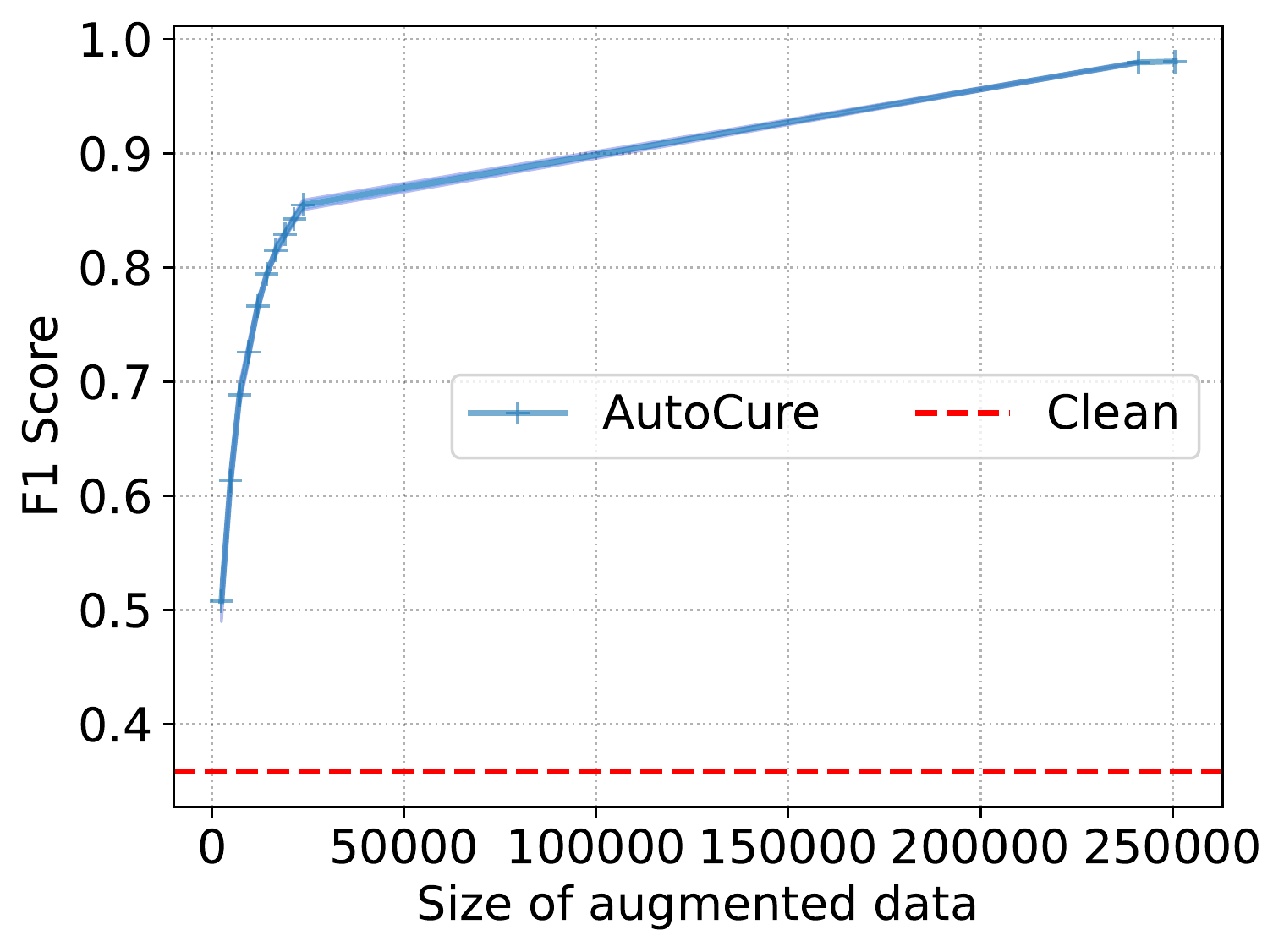}} \hfill
	\subfloat[Breast Cancer]{\label{fig:bcancer_aug}\includegraphics[width=0.2\linewidth]{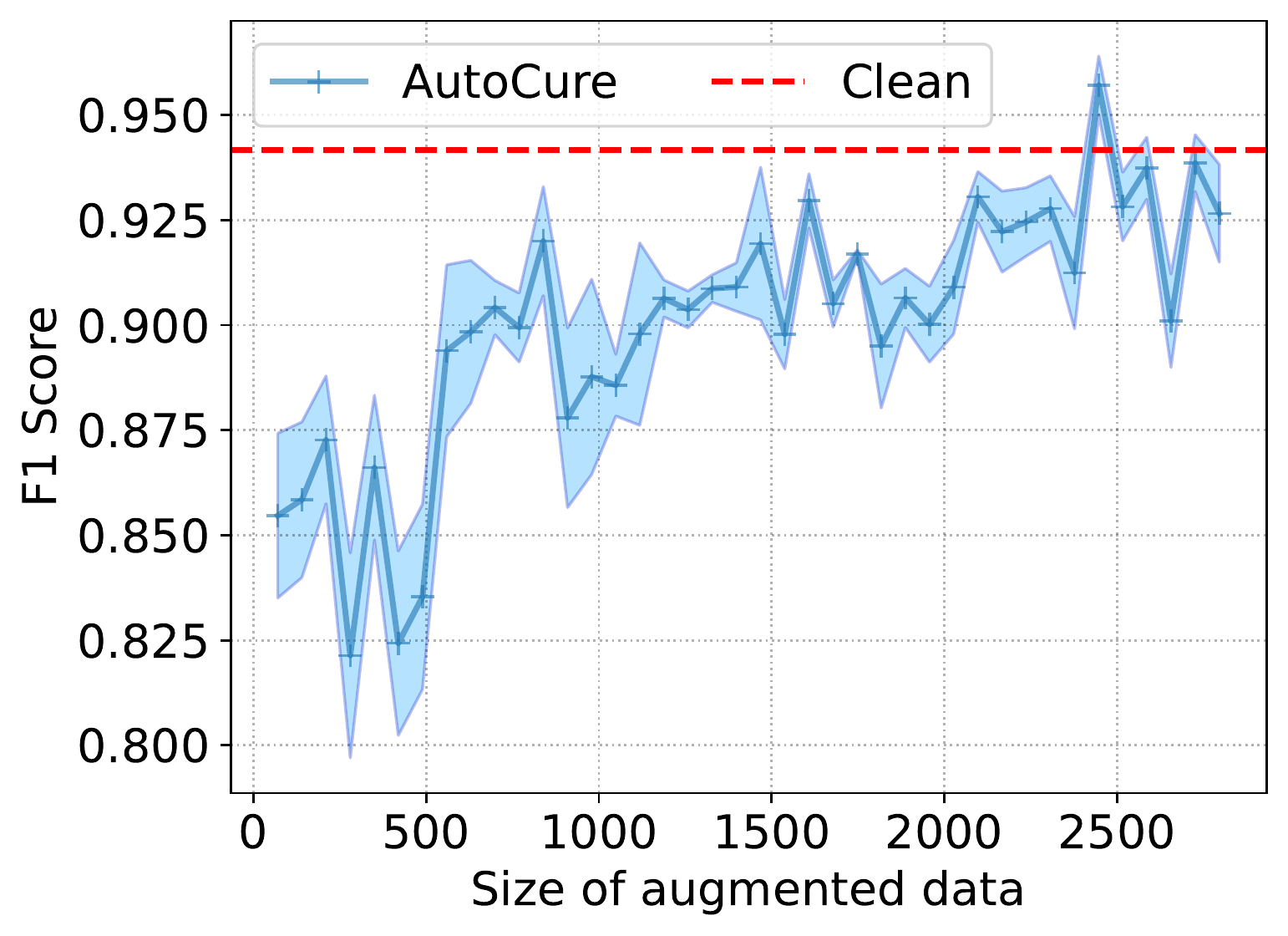}} \hfill
	\subfloat[Nasa]{\label{fig:nasa_aug}\includegraphics[width=0.2\linewidth]{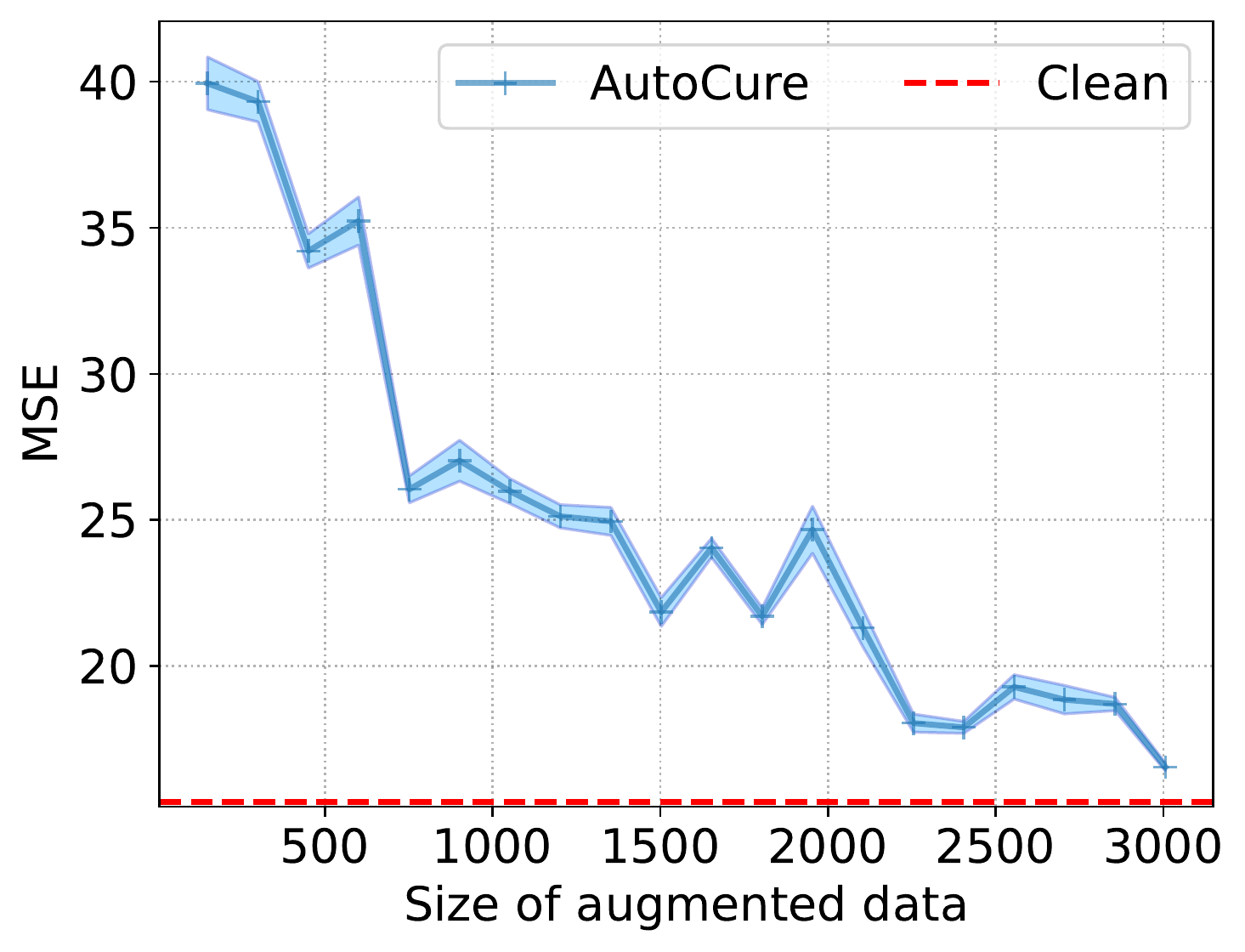}} \hfill
	\subfloat[Soil Moisture]{\label{fig:moisture_aug}\includegraphics[width=0.2\linewidth]{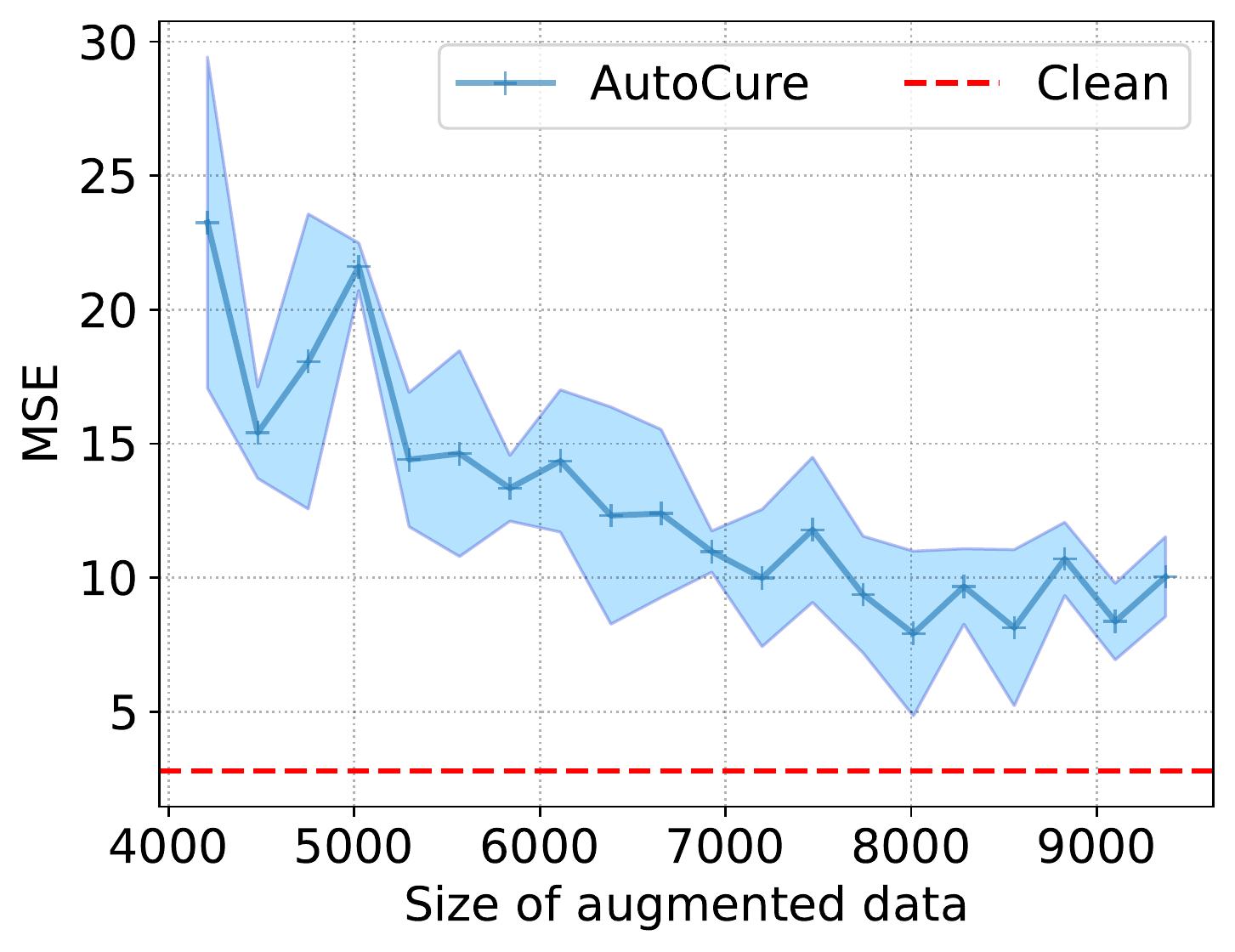}} \hfill
	\caption{Modeling accuracy of \PaperAcronym for different sizes of the augmented clean data}
	\label{fig:augmentation_figs}
\end{figure*}

\begin{figure*}[t]
	\centering
	\subfloat[Adult]{\label{fig:adult_aug_time}\includegraphics[width=0.2\linewidth]{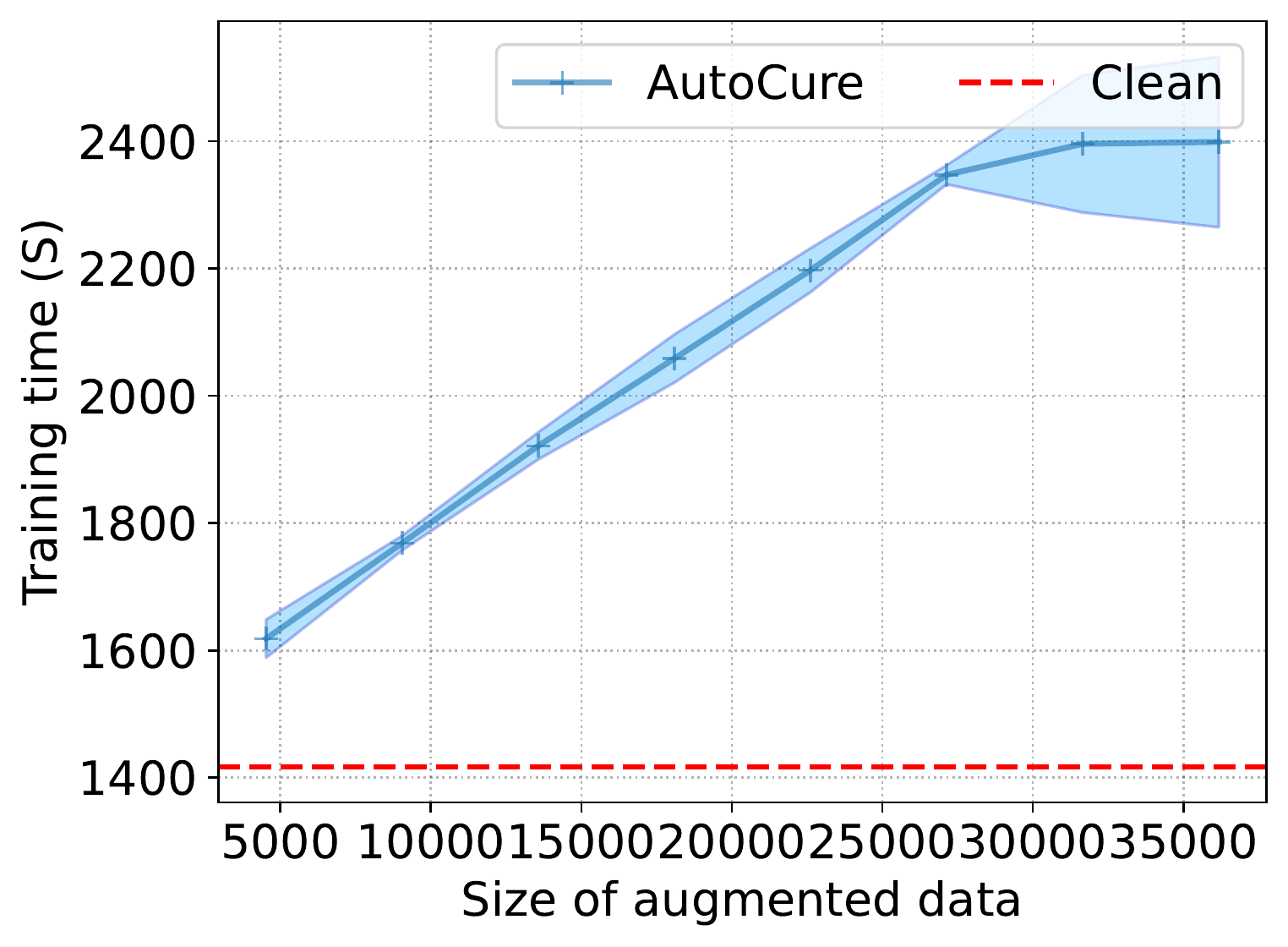}} \hfill
	\subfloat[Smart Factory]{\label{fig:factory_aug_time}\includegraphics[width=0.2\linewidth]{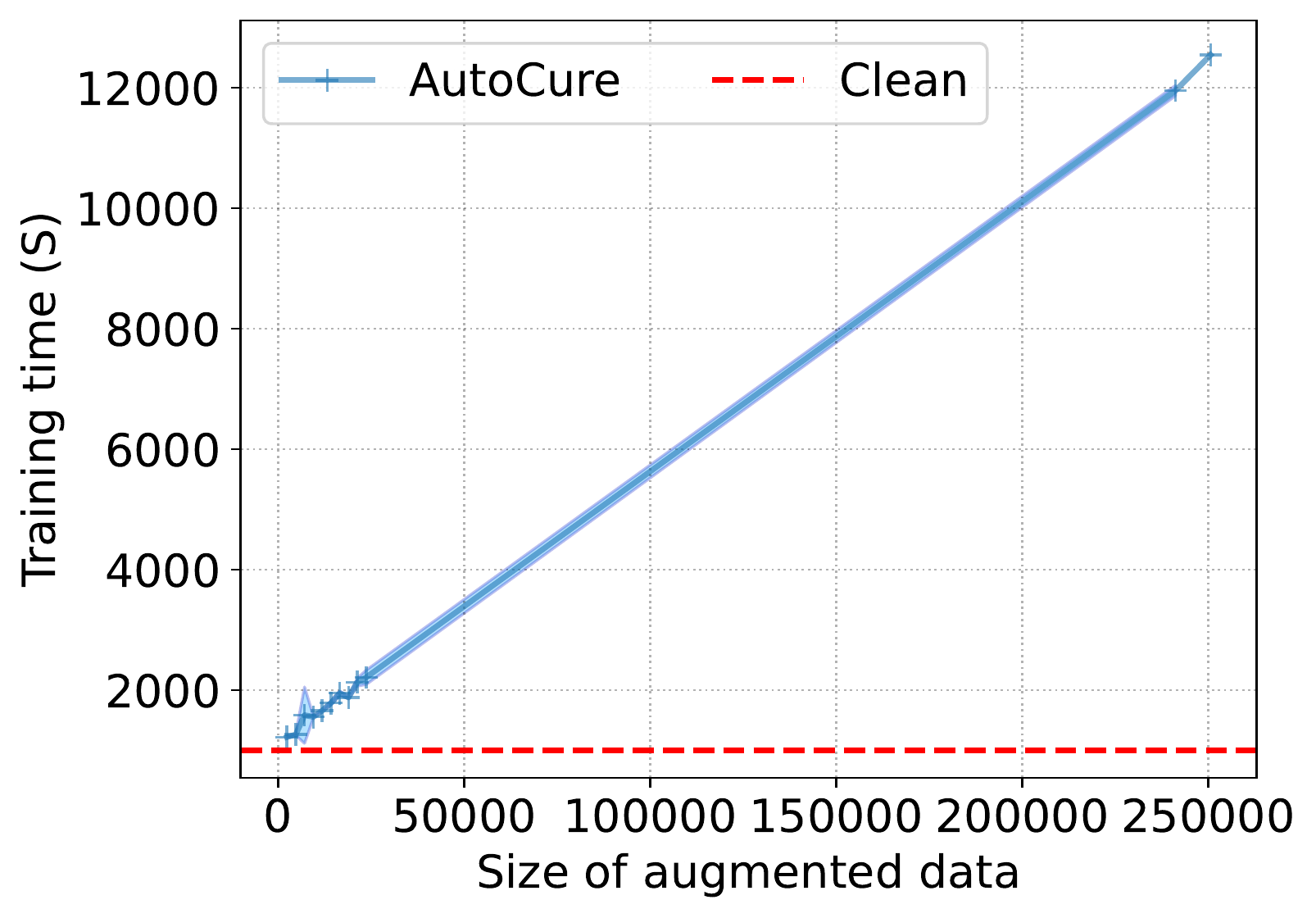}} \hfill
	\subfloat[Breast Cancer]{\label{fig:bcancer_aug_time}\includegraphics[width=0.2\linewidth]{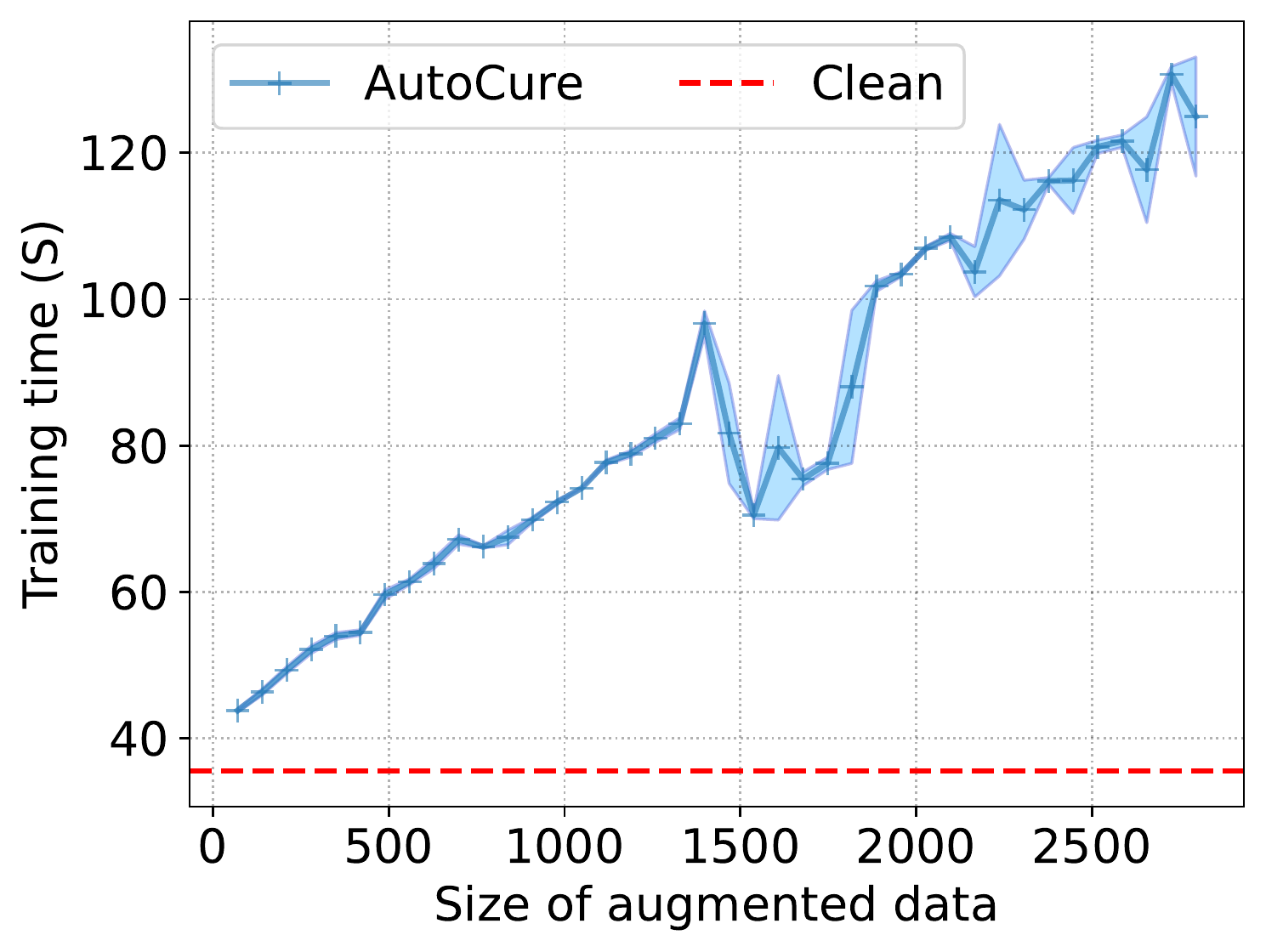}} \hfill
	\subfloat[Nasa]{\label{fig:nasa_aug_time}\includegraphics[width=0.2\linewidth]{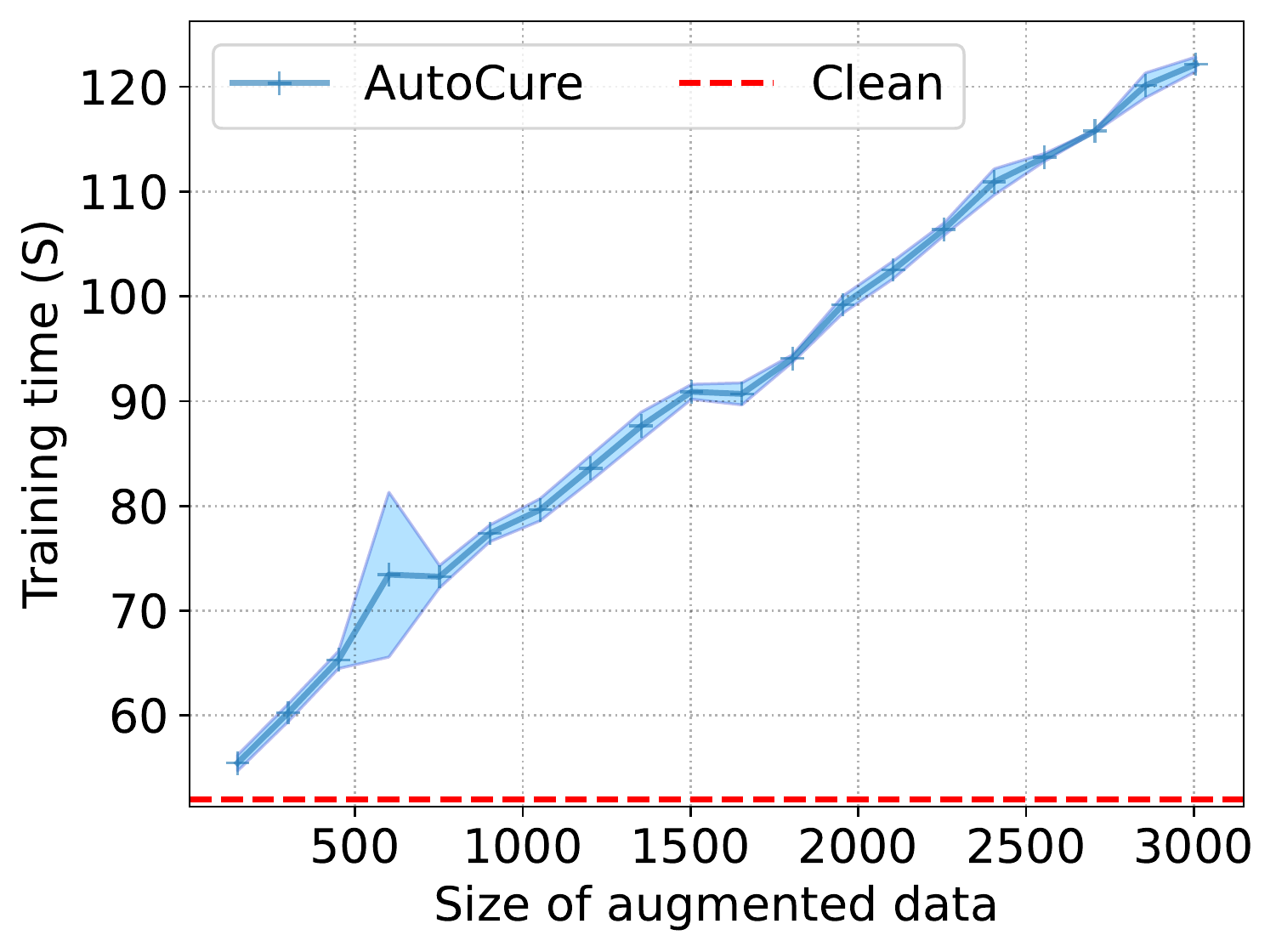}} \hfill
	\subfloat[Soil Moisture]{\label{fig:moisture_aug_time}\includegraphics[width=0.2\linewidth]{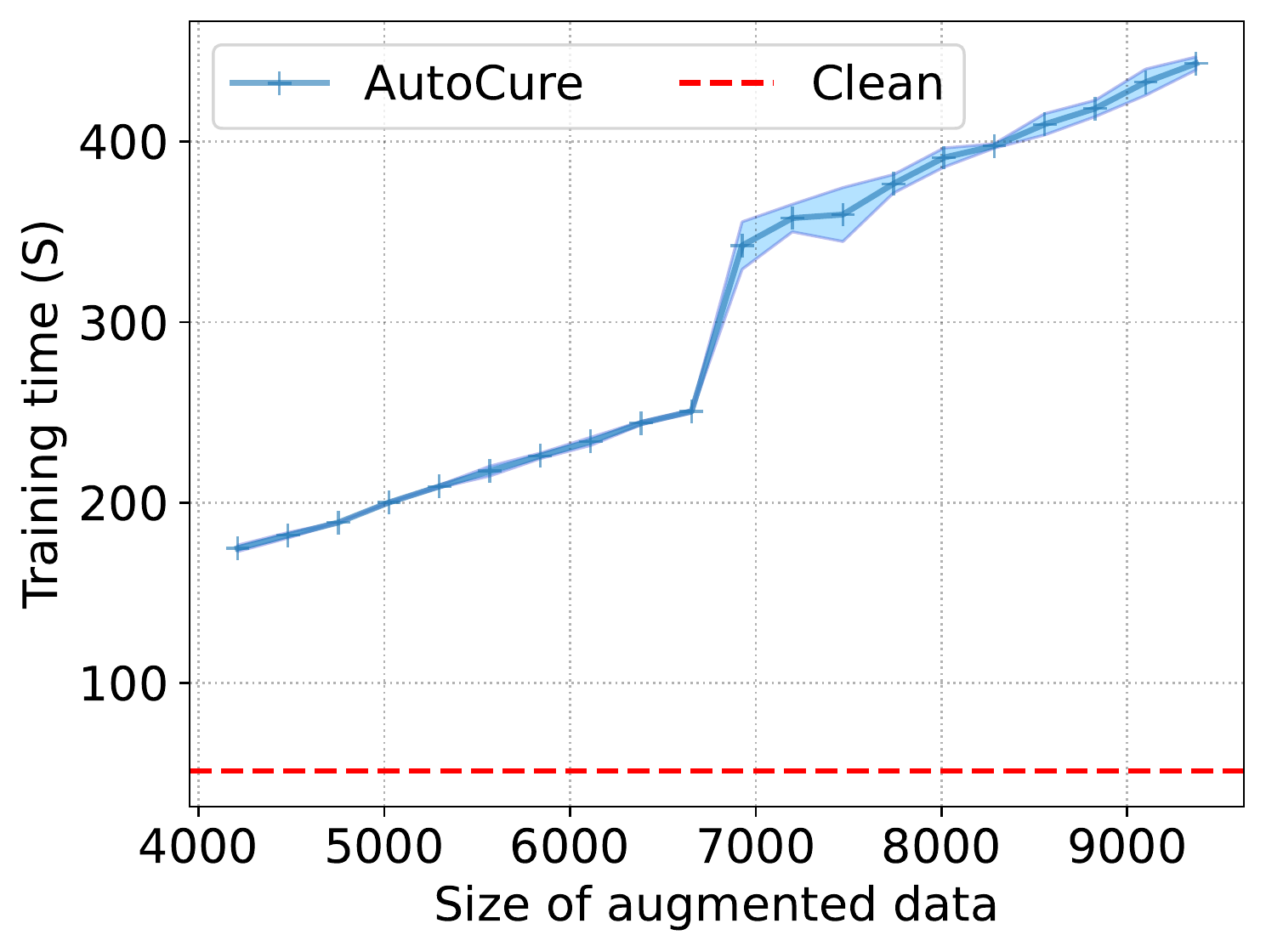}} \hfill
	\caption{Training time of \PaperAcronym for different sizes of the augmented clean data}
	\label{fig:augmentation_runtime_figs}
\end{figure*}

%
\begin{table}[tbph]
	\centering
	\ra{1}
	\caption{Average training time of compared methods}
	\label{tab:time}
	\resizebox{\columnwidth}{!}{%
		\begin{tabular}{@{}lcccccc@{}}
			\toprule[1.1pt]
			& \textbf{Adult} & \textbf{Cancer} & \textbf{Nasa} & \textbf{Moisture} & \textbf{Factory} & \textbf{Housing} \\ \midrule
			\textbf{Clean}    & 1416.8 & 35.5  & 51.9  & 51.4 & 996 & 43.5 \\ \midrule
			\textbf{B2}       & 1452.6 & 35.7  & 65.4  & 55.9 & 1265.8 & 45.65 \\
			\textbf{B3}       & 1467.2 & 41.4  & 63.2  & 52.6 & 1138.7 & 46.45 \\
			\textbf{E2}       & 1488.9 & 41.8  & 66.5  & 55.9 & 1138.2 & 47.08 \\
			\textbf{E3}       & 1487.6 & 37.1  & 54    & 51.6 & 1095.3 & 47.8 \\
			\textbf{H2}       & 1487.4 & 42.65 & 65.77 & 55.3 & 1152.4 & 45.83 \\
			\textbf{H3}       & 1495.8 & 33    & 53.89 & 55.4 & 1023.6 & 46.74 \\
			\textbf{K2}       & 1423.4 & 43.9  & 66.1  & 50.7 & 1105.7 & 42.85 \\
			\textbf{K3}       & 1370.5 & 38.1  & 63    & 46.3 & 1097.1 & 40.24 \\
			\textbf{R2}       & 1488.6 & 43.9  & 64.8  & 56.3 & 946.1 & 47.21 \\
			\textbf{R3}       & 1439.6 & 41.1  & 64.8  & 53.1 & 1166.9 & 45.61 \\
			\textbf{Combined} & 14601  & 398.4 & 627.5 & 533.1 & 11129.8 & 455.46 \\ \midrule
			\textbf{\PaperAcronym} & \textbf{1595.2} & \textbf{233.7} & \textbf{183.1} & \textbf{300.9} & \textbf{1352} & \textbf{300.6} \\ \bottomrule[1.1pt]
		\end{tabular}%
	}
\end{table}
%

\paragraph{Augmentation Size} In this set of experiments, we estimate, for each data set, the impact of the size of the augmented clean data on the predictive performance and the training time. A general remark from the experiments is that increasing the amount of augmented data broadly improves the predictive performance at the expense of increasing the training time. For instance, Figure~\ref{fig:adult_aug} shows the F1 score of the neural network trained on a clean data version of the Adult data set (cf. the red dashed line) and multiple versions curated by \PaperAcronym with different sizes of the augmented clean data. The figure shows that only a small amount of augmented data (circa 1\% of the size of the Adult data set) is sufficient to achieve the same performance as the clean data set. Similar results have been obtained in case of the Smart Factory data set, where 1\% of the size of the clean data set is sufficient to reach the quality of the clean data set.

For smaller data sets, the experiments showed that more clean data instances have to be generated. In the case of the Breast Cancer data set, Figure~\ref{fig:bcancer_aug} demonstrates that \PaperAcronym requires more clean instances (at least by 375\% relative to the size of the original data set) to achieve a similar accuracy as the clean data set. For the Nasa data set, \PaperAcronym achieves a comparable performance to that of the clean data set when the size of the generated clean data is large enough (circa 200\% of the size of the original data set) to minimize the impact of erroneous instances. Similar results have been obtained for the Soil Moisture data set (cf. Figure~\ref{fig:moisture_aug}). 

Figure~\ref{fig:augmentation_runtime_figs} depicts the training times of different data sets curated by \PaperAcronym. For example, Figure~\ref{fig:adult_aug_time} illustrates the training time of the clean version (cf. the red dashed line) and several versions curated by \PaperAcronym. As the figure depicts, \PaperAcronym requires circa three minutes more than it can be needed by the clean version of the data set. In case of the Smart Factory data set, \PaperAcronym requires slightly more time (at most by 18.4\%) to satisfy the quality requirements. For the Breast Cancer data set, \PaperAcronym requires an additional 100 seconds to achieve a similar accuracy as the clean data (cf. Figure~\ref{fig:bcancer_aug_time}). Along a similar line, an additional 80 seconds are needed by \PaperAcronym in case of the Nasa data set (cf. Figure~\ref{fig:nasa_aug_time}). For the Soil Moisture data set, due to the associated regression task, \PaperAcronym has to generate numerous clean instances. As a result, \PaperAcronym requires additional training time of circa 400 seconds to reach the performance upper-bound.  

\begin{figure}[tbph]
	\centering
	\subfloat[Nasa]{\label{fig:nasa_er}\includegraphics[width=0.33\columnwidth]{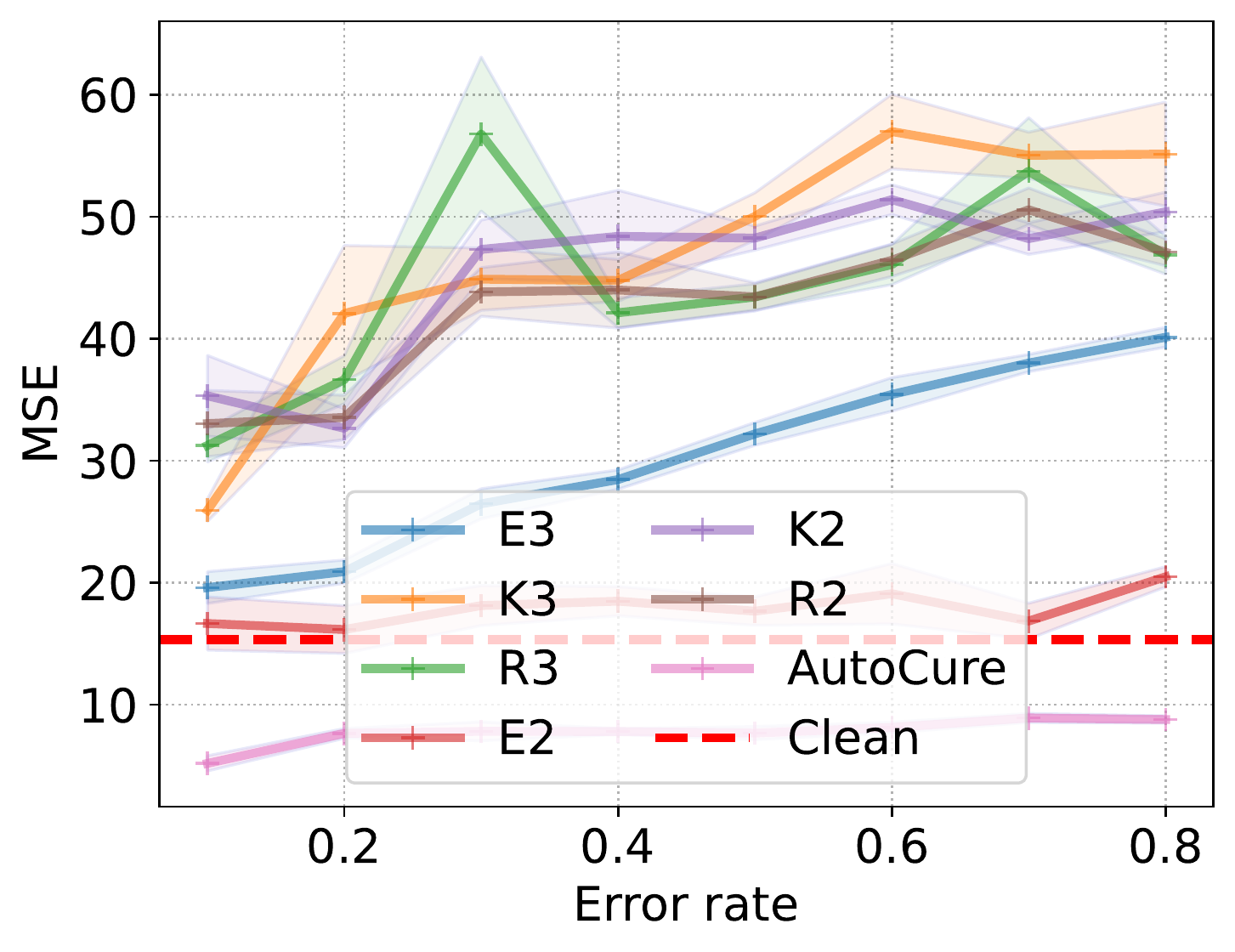}} \hfill
	\subfloat[Housing]{\label{fig:housing_er}\includegraphics[width=0.33\columnwidth]{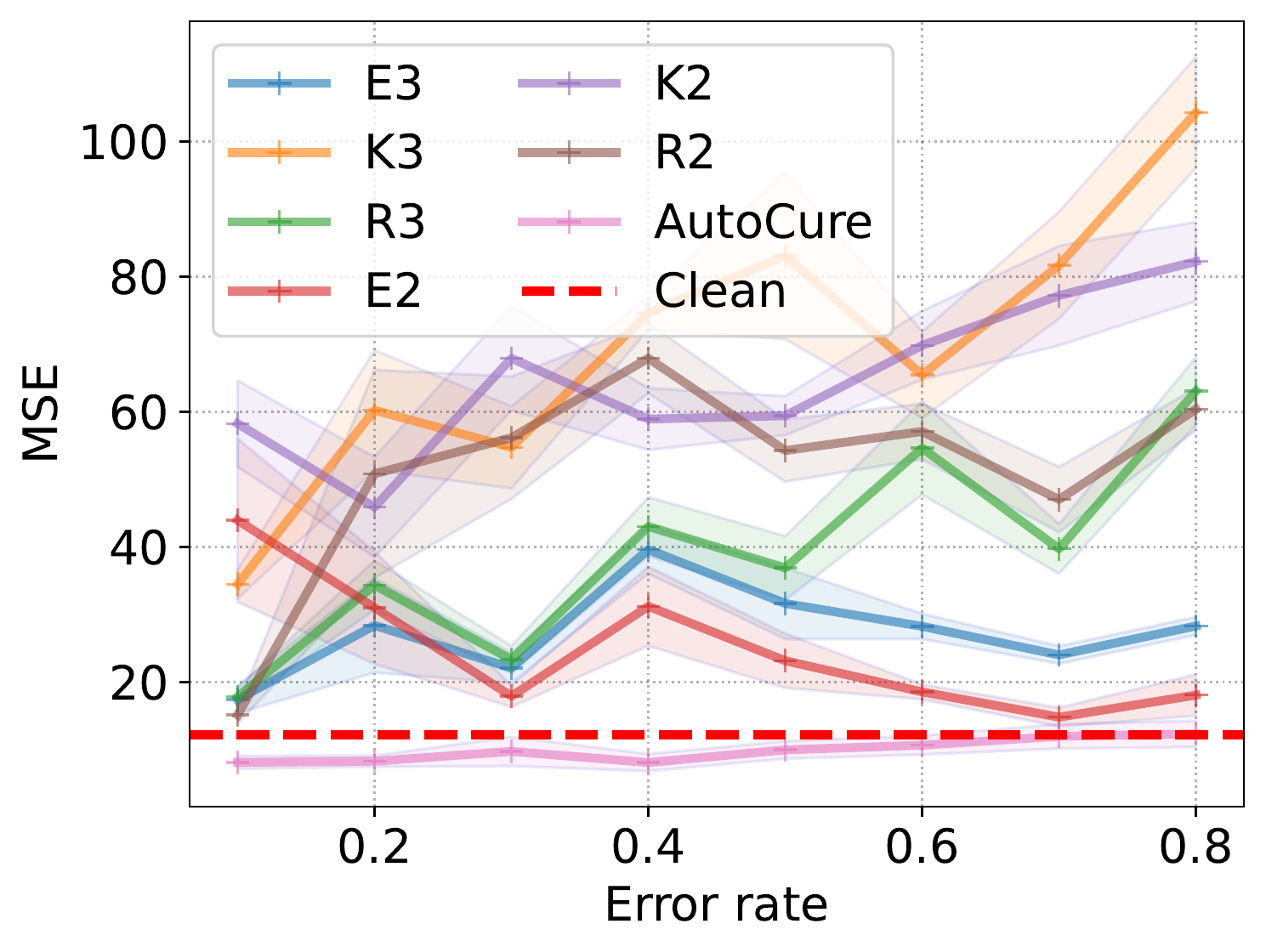}} \hfill
	\subfloat[Breast Cancer]{\label{fig:bcancer_er}\includegraphics[width=0.33\columnwidth]{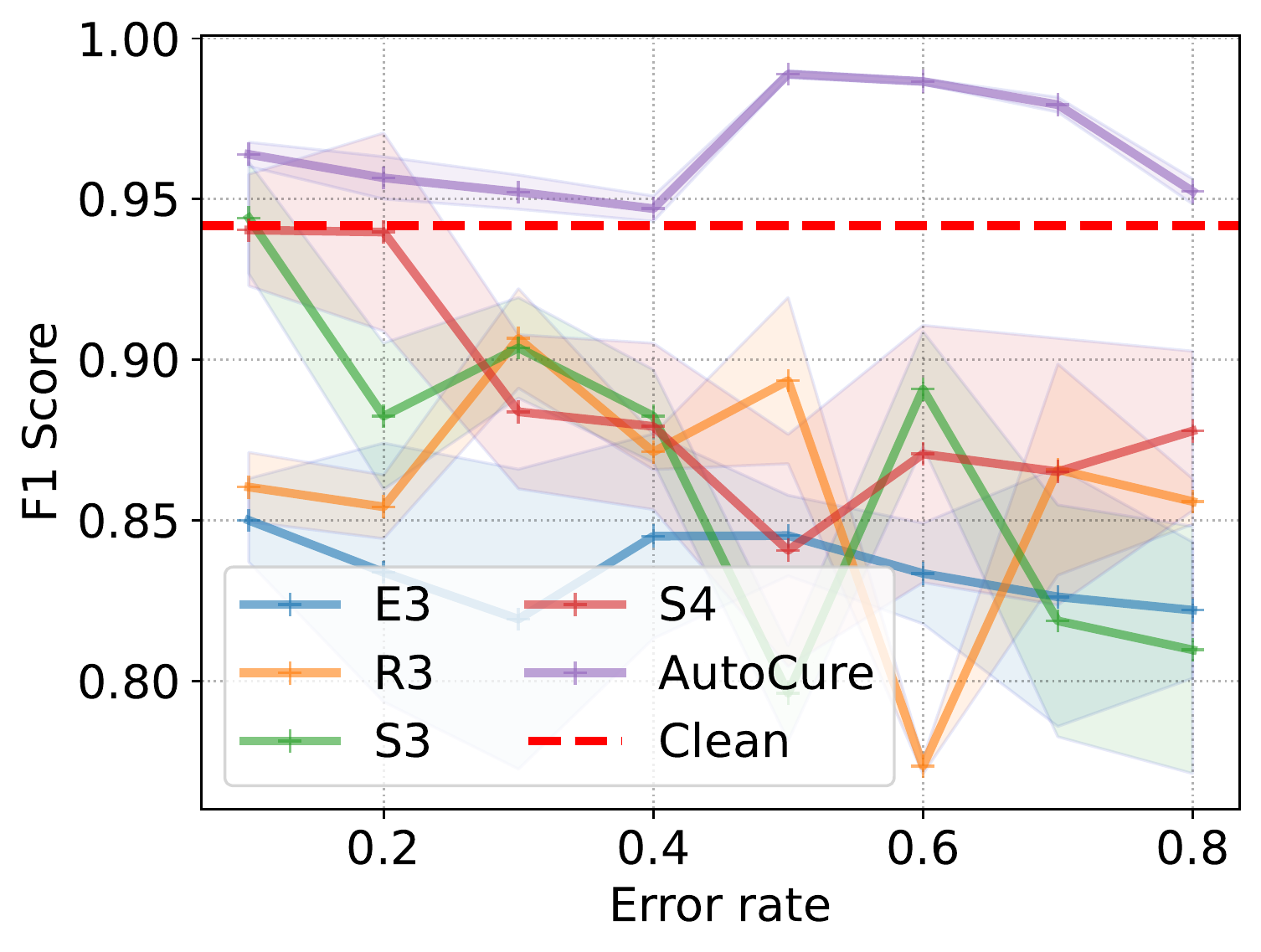}} \hfill
\caption{Impact of error rate on the predictive accuracy}
\label{fig:er_figs}
\end{figure}

\begin{figure}[tbph]
	\centering
	\subfloat[Nasa]{\label{fig:nasa_er_time}\includegraphics[width=0.33\columnwidth]{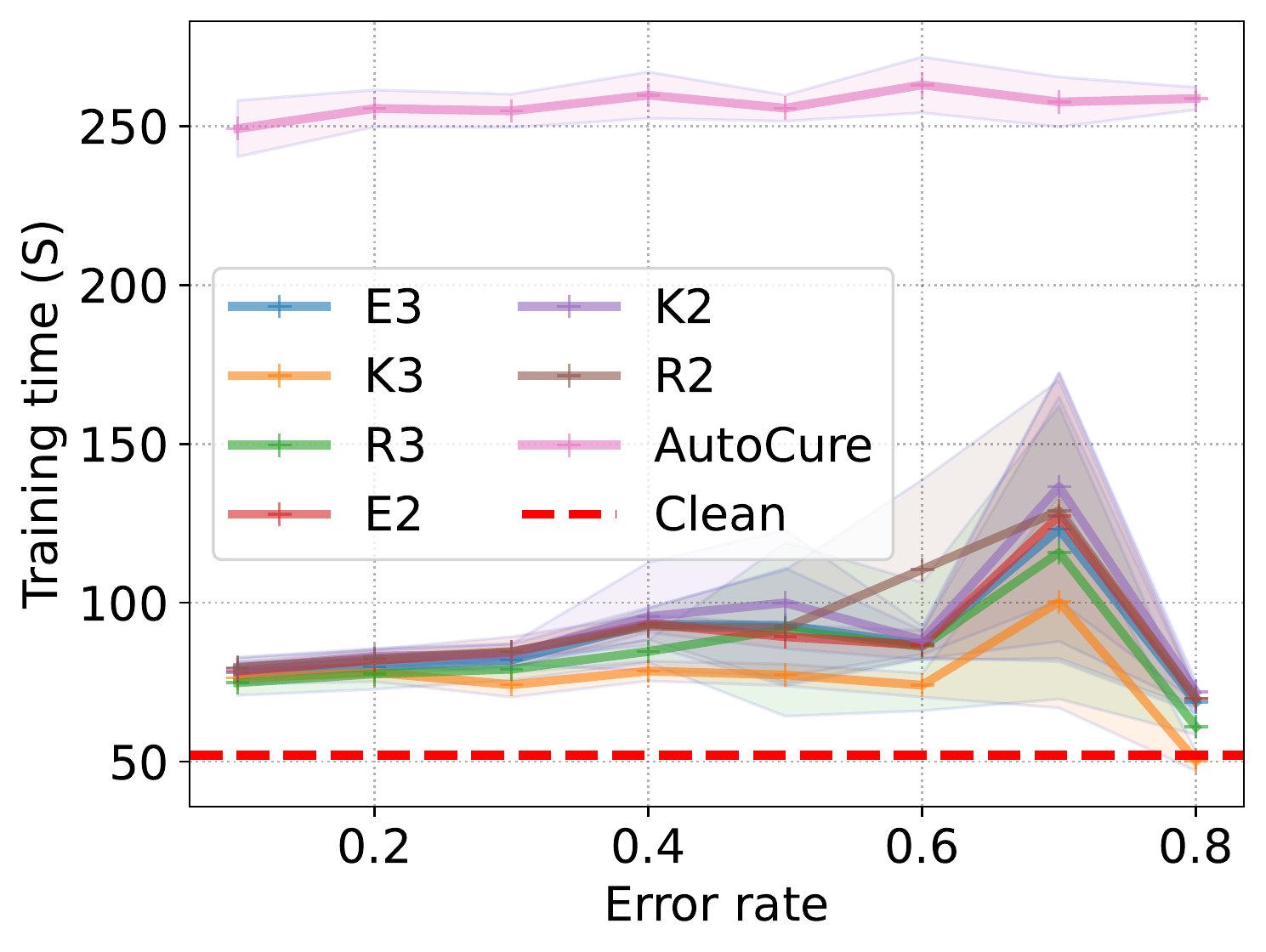}} \hfill
	\subfloat[Housing]{\label{fig:housing_er_time}\includegraphics[width=0.33\columnwidth]{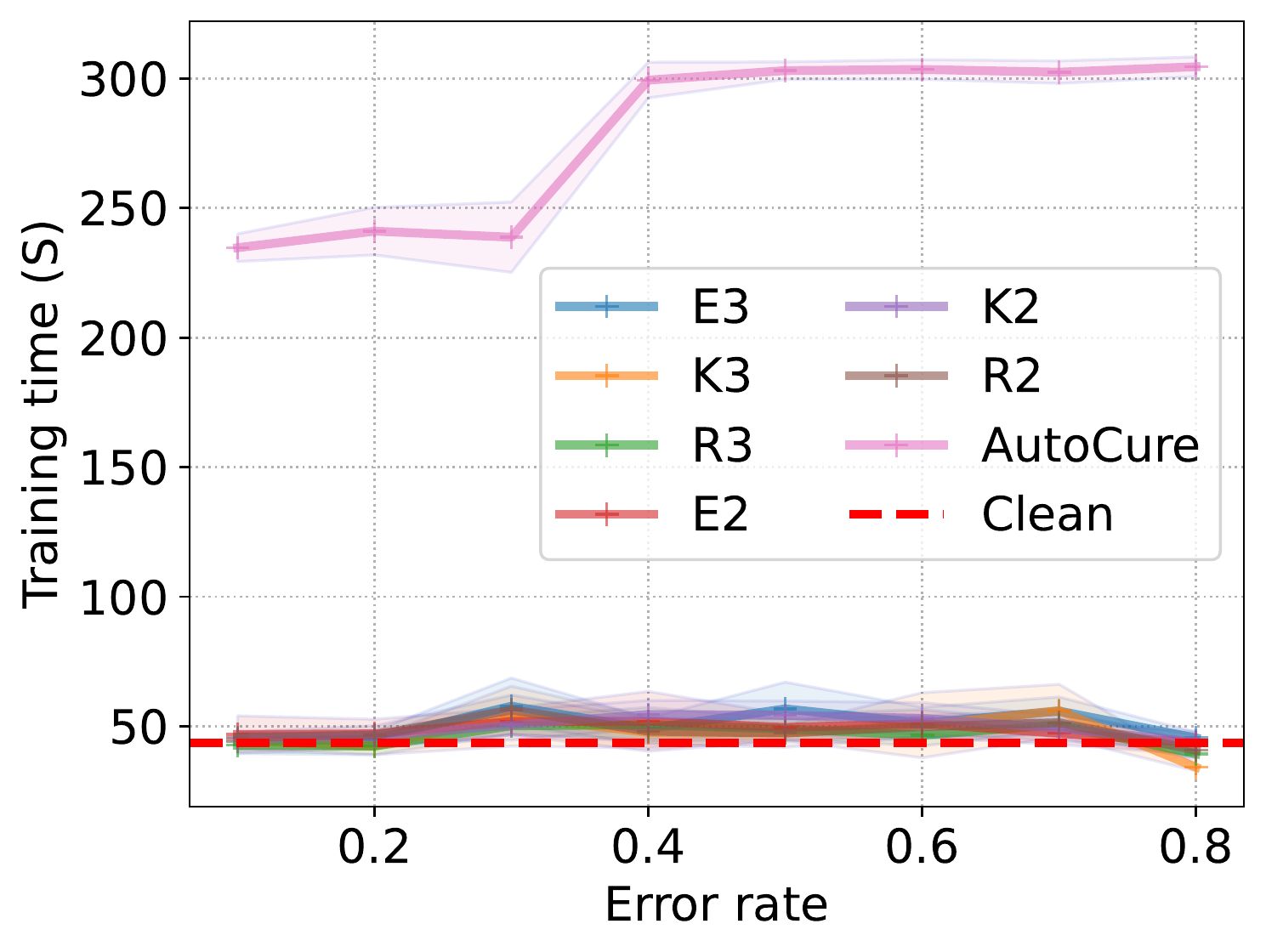}} \hfill
	\subfloat[Breast Cancer]{\label{fig:bcancer_er_time}\includegraphics[width=0.33\columnwidth]{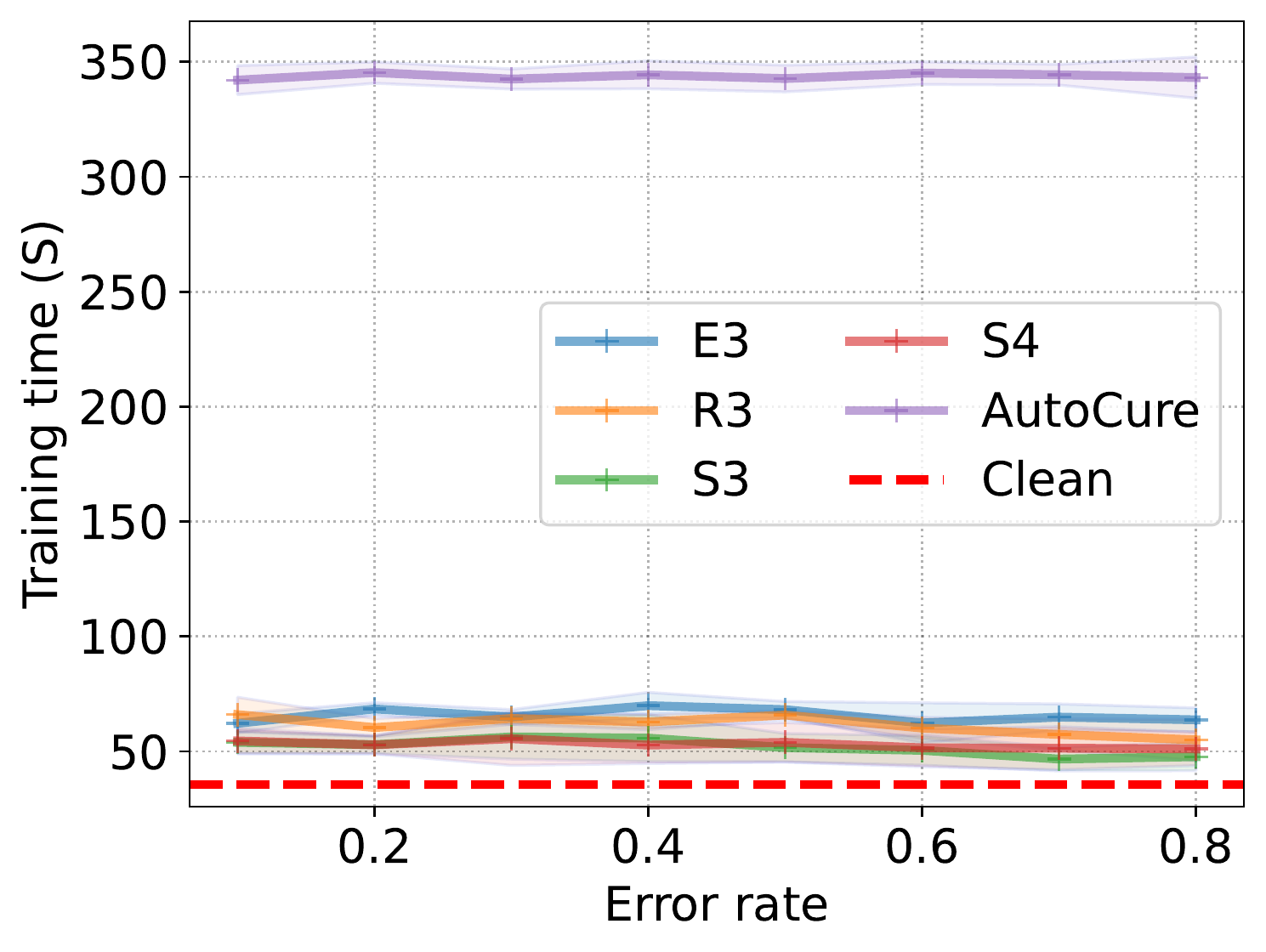}} \hfill
	\caption{Impact of error rate on the training time}
	\label{fig:er_timing_figs}
\end{figure}

\paragraph{Error Rate} In this set of experiments, we assess the impact of the error rate on the performance of \PaperAcronym and a set of baseline methods. We fixed the number of generated clean instances and gradually increased the amount of injected errors (i.e., missing values and numerical outliers). Figure~\ref{fig:er_figs} depicts the predictive accuracy of the neural network, in terms of MSE for regression tasks and F1 score for classification tasks, while being trained on the clean data set and the curated data sets using \PaperAcronym and the baseline methods. In Figure~\ref{fig:nasa_er}, \PaperAcronym shows that it is mostly agnostic to the amount of errors in the Nasa data set. Even with a fixed number of generated clean instances, the curve of \PaperAcronym is slightly increased, which is still below the red dashed line of the clean data set. Conversely, all traditional data curation methods poorly perform with increasing the error rate. Figures~\ref{fig:housing_er} and \ref{fig:bcancer_er}) show similar results for the Housing and Breast Cancer data sets. Figures~\ref{fig:nasa_er_time}-\ref{fig:bcancer_er_time} shows the training time of the neural network while increasing the error rate. Obviously, increasing the error rate has almost no influence on the training time of all examined data sets (except some spikes which rarely occur at certain error rates).


\begin{figure}[tbph]
	\centering
	\subfloat[Smart Factory]{\label{fig:factory_lc}\includegraphics[width=0.33\columnwidth]{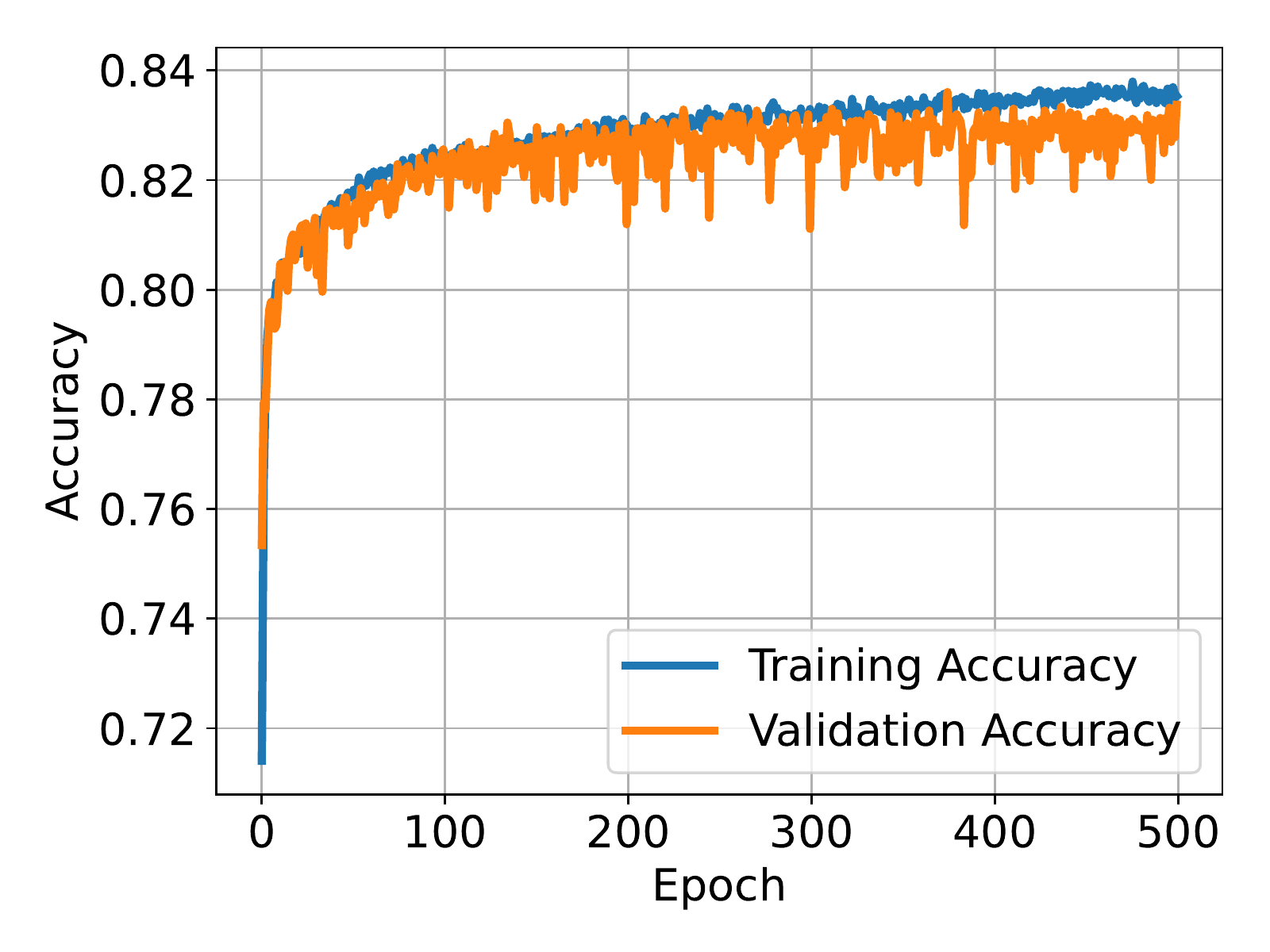}} \hfill
	\subfloat[Adult]{\label{fig:adult_lc}\includegraphics[width=0.33\columnwidth]{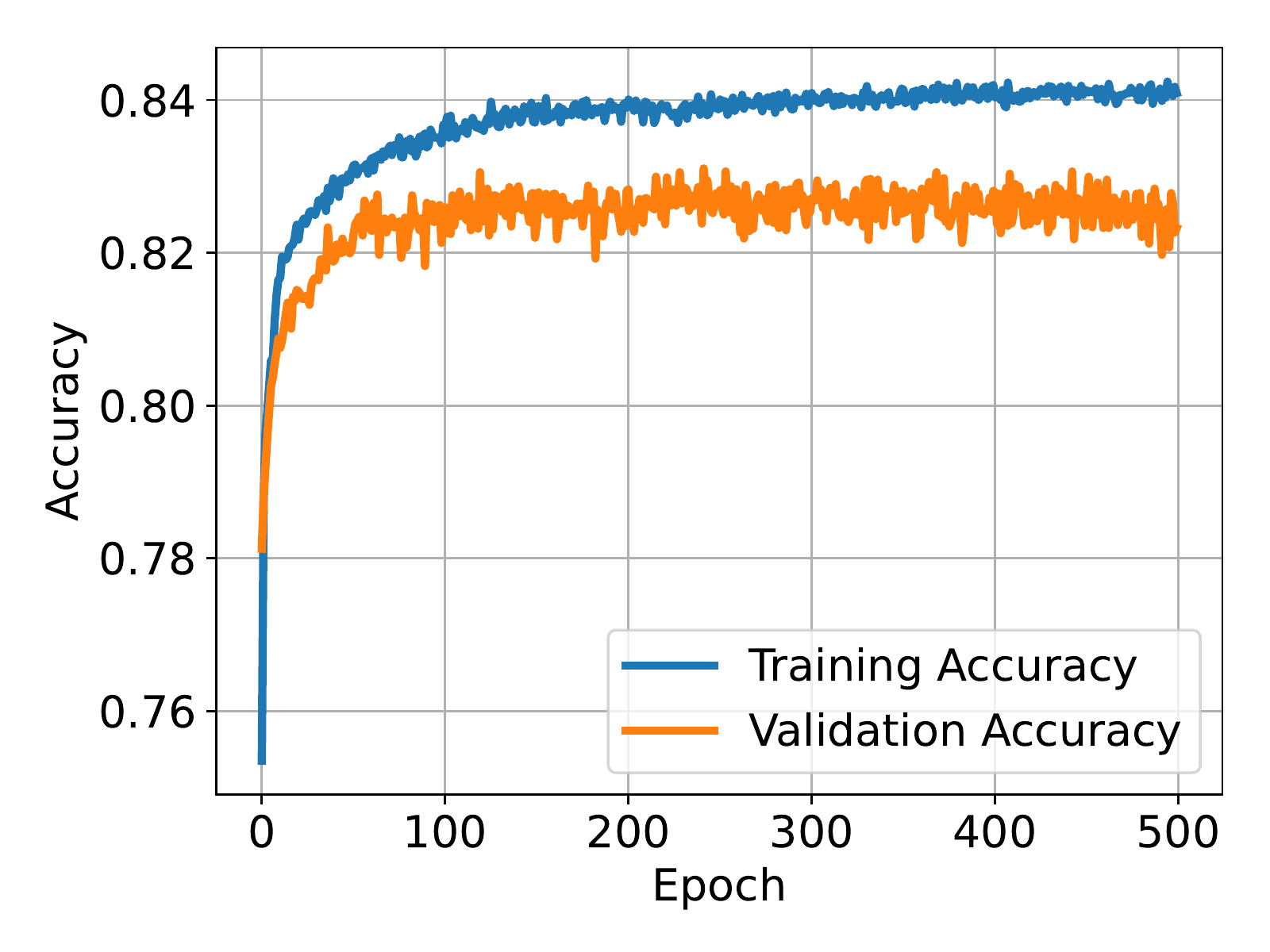}} \hfill
	\subfloat[Breast Cancer]{\label{fig:bcancer_lc}\includegraphics[width=0.33\columnwidth]{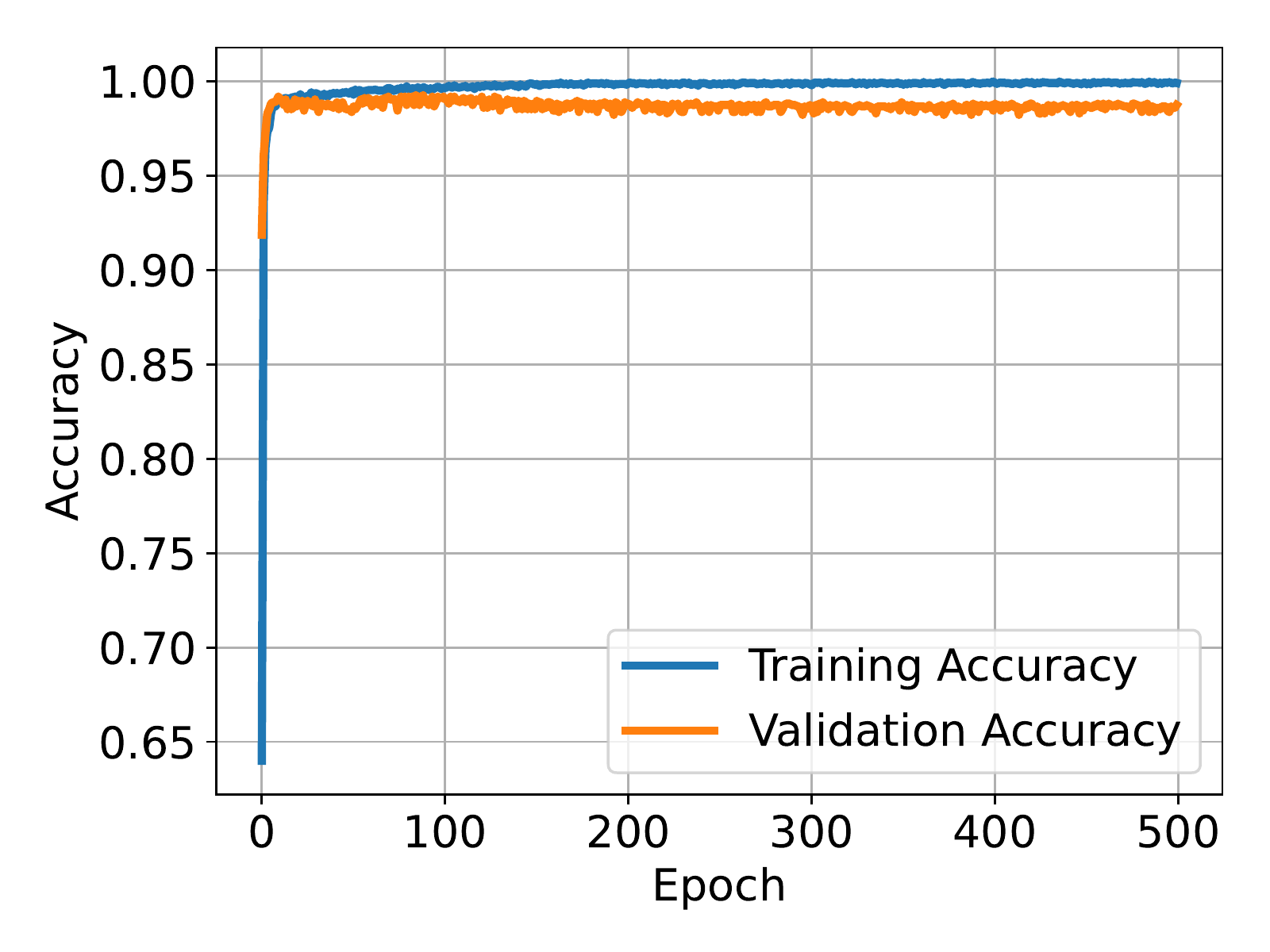}} \hfill
	\subfloat[Nasa]{\label{fig:nasa_lc}\includegraphics[width=0.33\columnwidth]{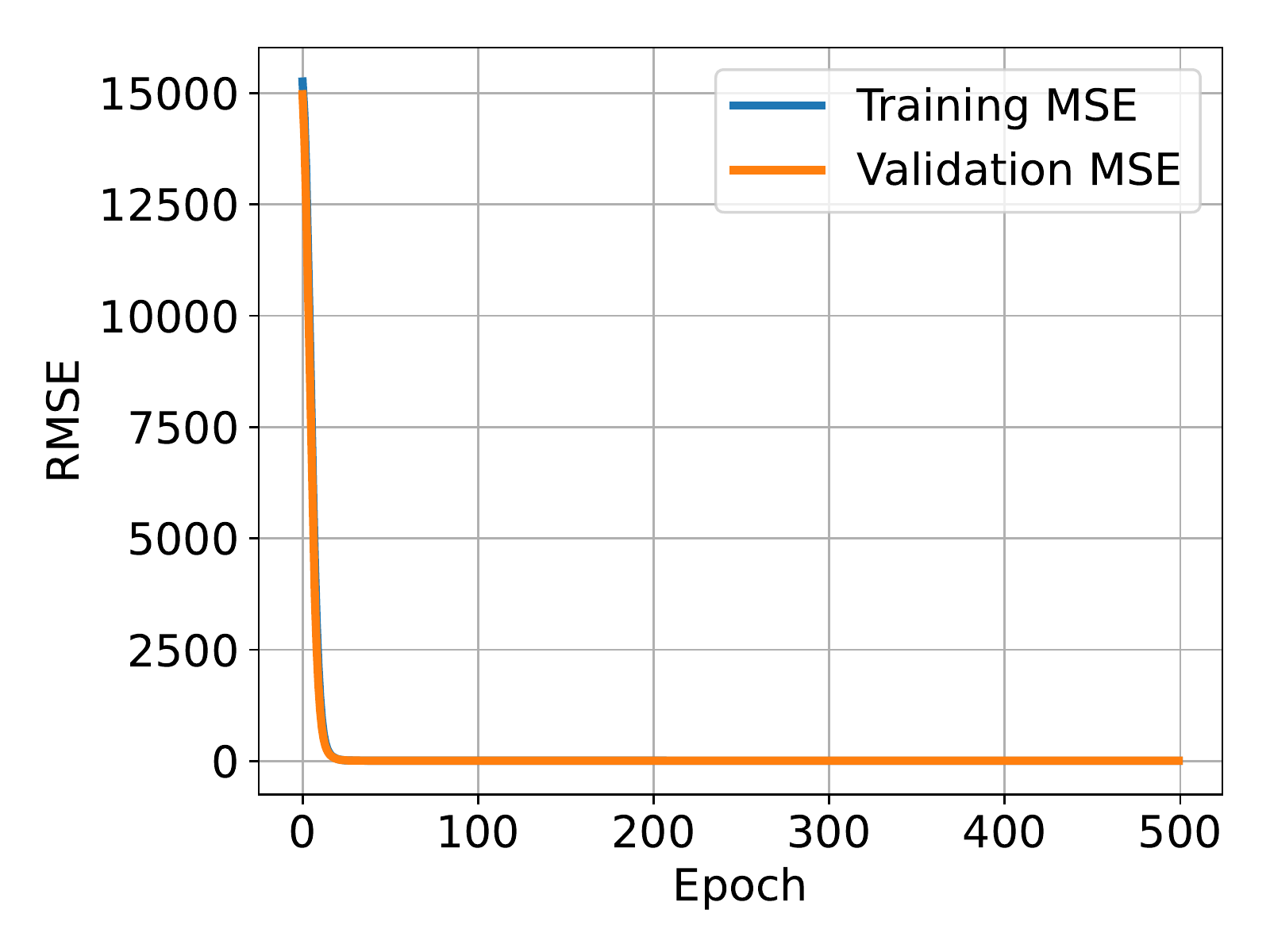}} \hfill
	\subfloat[Soil Moisture]{\label{fig:moisture_lc}\includegraphics[width=0.33\columnwidth]{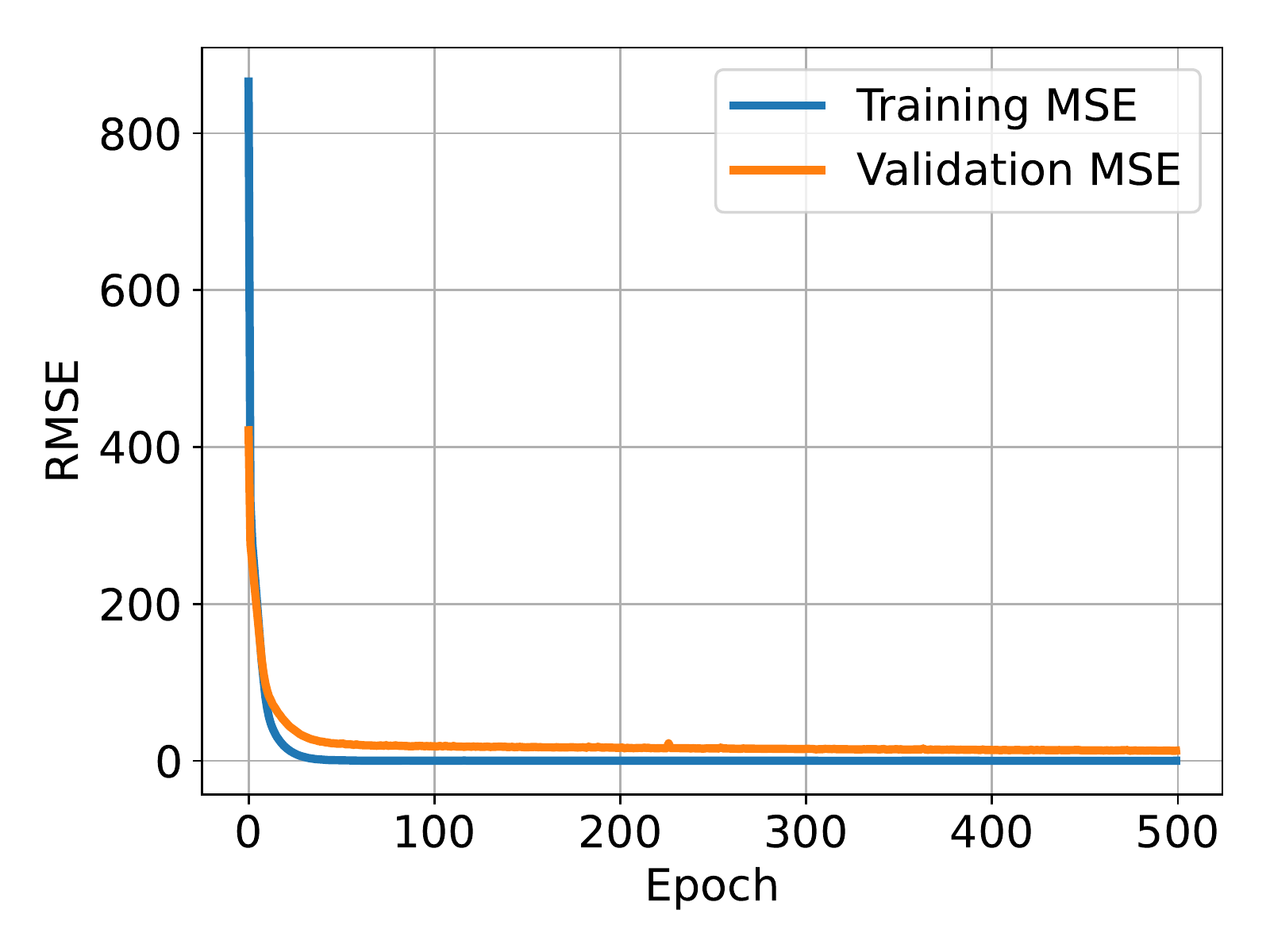}} \hfill
	\subfloat[Housing]{\label{fig:housing_lc}\includegraphics[width=0.33\columnwidth]{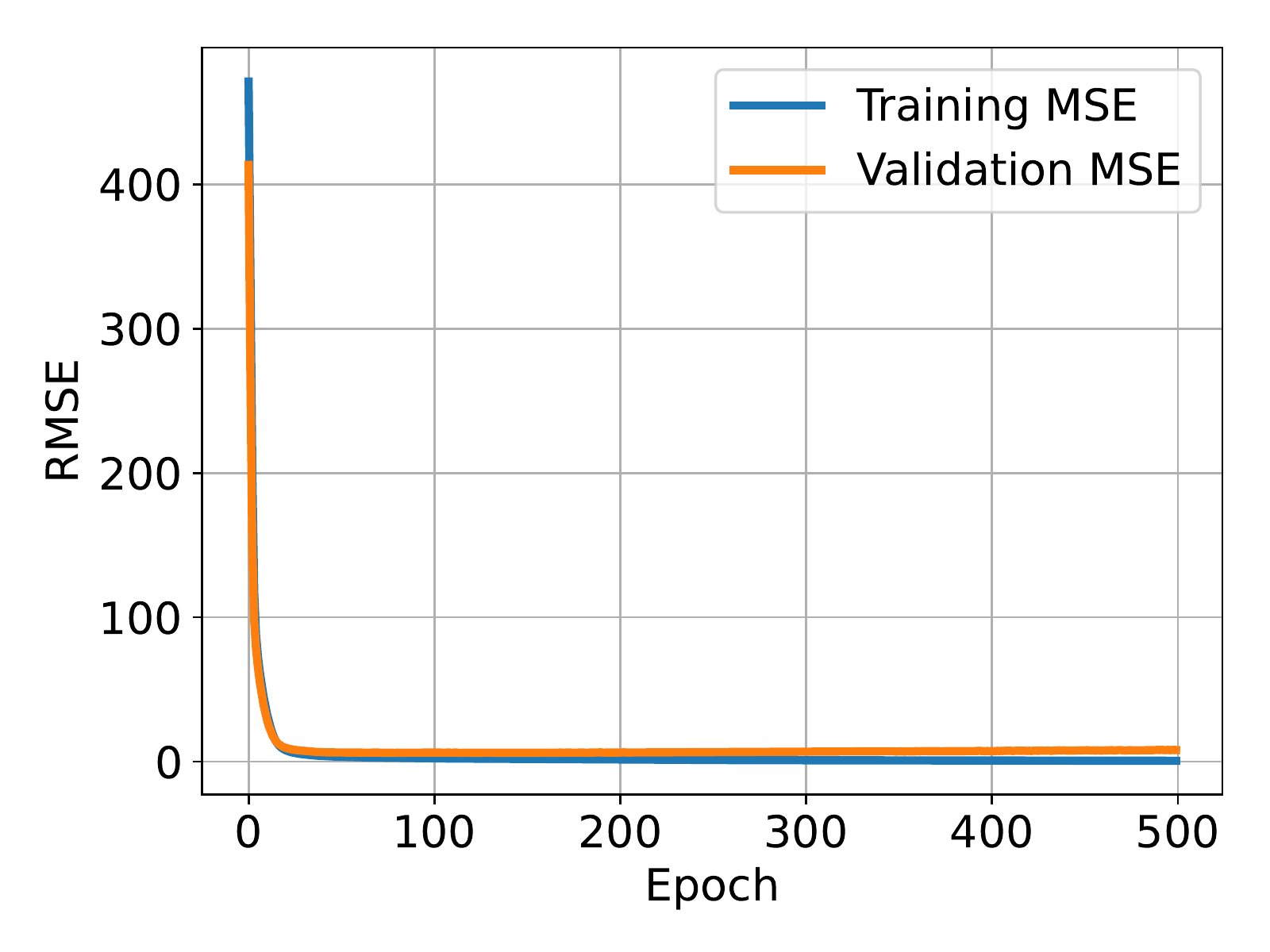}} \hfill
	\caption{Learning curves of the neural networks trained using \PaperAcronym}
	\label{fig:lc_figs}
\end{figure}

\paragraph{Overfitting} The results, discussed above, show that \PaperAcronym broadly improves the predictive performance while requiring no repair operations. However, such a high performance may be a result of overfitting. Therefore, in this set of experiments, we examine the generated neural network models, while using \PaperAcronym, for overfitting. Figure~\ref{fig:lc_figs} depicts the learning curves, with respect to the number of epochs and the predictive performance, for several data sets. For instance, in the case of the Smart Factory data set, Figure~\ref{fig:factory_lc} demonstrates that the model performance, in terms of the binary accuracy, is growing over time till it reaches a plateau. Obviously, both curves for the training and the validation sets are close to each other. This result has been repeated for all other data sets (cf. Figures~\ref{fig:adult_lc}-\ref{fig:housing_lc}). Accordingly, the high performance of \PaperAcronym, depicted in Figure~\ref{fig:radar_figs}, is not a result of overfitting, and it mainly occurs due to the high density of clean data instances. 

\subsection{Results Summary}
%
In this section, we highlight the main findings obtained in this work. The first finding revolves around the performance consistency. Specifically, our results revealed that some baseline methods may achieve similar or even better performance than \PaperAcronym (e.g., accuracy of the models trained on data curated by E2, E3, and H2 in Figure~\ref{fig:factory}). Nevertheless, it is important to mention that these results are broadly data set dependent, i.e., the performance may vary from one data set to another. Moreover, multiple experiments have to be carried out to identify a well-suited curation strategy for each data set. Alternatively, the results showed that \PaperAcronym shows a consistency of the results over different data sets. For all examined data sets, \PaperAcronym managed to improve the quality of the data set through emphasizing the clean fractions.  

Another important finding is that small data sets usually require the generation of more clean data instances (cf. Figure~\ref{fig:bcancer_aug}). This behavior of \PaperAcronym is caused since the percentage of original clean instances, in small data sets, is typically small. Accordingly, \PaperAcronym generates a large number of clean instances to satisfy the quality requirements. This finding also applies to data sets with associated regression tasks, e.g., Housing, Soil Moisture, and Nasa (cf. Figures~\ref{fig:nasa_aug} and \ref{fig:moisture_aug}). In fact, regression models are generally more sensitive to data errors than classification tasks, as found in our experiments. For large data sets, e.g., Adult and Smart Factory, \PaperAcronym requires much less generated data to achieve a similar accuracy as the clean data set. Furthermore, the results also showed that \PaperAcronym, in some cases, increases the training time due to increasing the number of training instances (cf. Figures~\ref{fig:bcancer_aug_time}, \ref{fig:nasa_aug_time} and \ref{fig:moisture_aug_time}). Finally, the results confirmed that \PaperAcronym can perform well even with high error rates, as shown in Figure~\ref{fig:er_figs}.
    
%% file: sections/related_work.tex
\section{related work}
\label{sec:related_work}
In this section, we report on the state-of-the-art techniques and tools relevant to the data curation problem. In fact, there exist plenty of data curation techniques and tools from academia and industry. For instance, HoloClean \cite{holoclean17} is an ML-agnostic data repair technique which infers the repair values via holistically employing multiple cleaning signals to build a probabilistic graph model. To repair pattern violations and inconsistencies, OpenRefine \cite{openrefine13} utilizes Google Refine Expression Language (GREL) as its native language to transform existing data or to create repair values. Similarly, BARAN \cite{baran20} is a holistic configuration-free ML-based method for repairing different error types. To this end, BARAN trains incrementally updatable models which leverage the value, the vicinity, and the domain contexts of data errors to propose correction candidates. To further increase the training data, BARAN exploits external sources, such as Wikipedia page revision history. 

In fact, the above presented techniques, such as HoloClean, OpenRefine, and BARAN, do not consider the requirements imposed by the downstream ML applications. They tend to improve the data quality regardless of where and how the data comes from or how the data will be consumed. Therefore, a new set of techniques and tools has been emerged which strives to jointly optimize the cleaning and modeling tasks. In other words, these ML-oriented methods focus on selecting the optimal repair candidates with the objective of improving the performance of specific predictive models. Accordingly, these methods assume the availability of repair candidates from other ML-agnostic methods. For instance, BoostClean \cite{boostclean17} deals with the error repair task as a statistical boosting problem. Specifically, it composes a set of weak learners into a strong learner. To generate the weak learners, BoostClean iteratively selects a pair of detection and repair methods, before applying them to the training set to derive a new model. 

ActiveClean \cite{activeclean16} is another ML-oriented method, principally employed for models with convex loss functions. It formulates the data cleaning task as a stochastic gradient descent problem. Initially, it trains a model on a dirty training set, where such a model is to be iteratively updated until reaching a global minimum. In each iteration, ActiveClean samples a set of records and then asks an oracle to clean them to shift the model along the steepest gradient. A similar work is CPClean \cite{cpclean20} which incrementally cleans a training set until it is certain that no more repairs can possibly change the model predictions. In fact, the ML-oriented methods do not introduce new data repair techniques, such as HoloClean and BARAN. Instead, they tend to select the already-existent repair candidates which may improve the predictive performance. In comparison to \PaperAcronym, such methods can be entirely avoided since no repair candidates have to be generated in our solution. Accordingly, \PaperAcronym can avoid the complexities of searching for the repair candidates and selecting the best subset from those candidates. Furthermore, the ML-oriented methods are usually tailored to specific optimization methods and ML models, e.g., ActiveClean limited to problems with convex loss functions. Alternatively, \PaperAcronym can be utilized with all ML tasks and optimization methods.

%% file: sections/conclusion.tex
\section{Conclusion \& Future Work}
\label{sec:conclusion}

In this paper, we introduce a novel method for dealing with erroneous data, referred to as \PaperAcronym. As alternative to traditional data curation tools, \PaperAcronym does not require the recognition of the detected error types, since it entirely ignores such noisy data instances. Instead, it deals with the clean instances which can lead to a better modeling performance, if more clean instances do exist. Moreover, \PaperAcronym avoids the complicated processes for searching the huge space of repair candidates and the process of selecting the most suitable candidates. As a proof of concept, \PaperAcronym has been evaluated using six data sets and over 28 combinations of data curation tools. The results showed that the data sets curated using \PaperAcronym can achieve a similar or even better performance as the clean data set. While \PaperAcronym delivers an outstanding performance over multiple data sets, it can be further improved through: (1) integrating it with a data valuation tool, to reduce the training time, and (2) dynamically adapting the number of instances to be augmented depending on the size of the dirty data set and the associated ML task.

